\documentclass[11pt,showpacs,amsmath,nofootinbib,superscriptaddress]{revtex4}
\usepackage{amssymb}
\usepackage[hypertex]{hyperref}
\usepackage{graphicx}
\usepackage{epstopdf}
\usepackage{bm}
\usepackage{epsfig}
\usepackage{graphics}
\usepackage{xspace}
\usepackage{amsfonts}
\usepackage{color}

\begin{document}


\title{Threshold resummation for the production of a color sextet (antitriplet) scalar at the LHC}

\vspace*{1cm}

\author{Yong Chuan Zhan}
\affiliation{Department of Physics and State Key Laboratory of Nuclear Physics and Technology, Peking University, Beijing, 100871, China}

\author{Ze Long Liu}
\affiliation{Department of Physics and State Key Laboratory of Nuclear Physics and Technology, Peking University, Beijing, 100871, China}

\author{Shi Ang Li}
\affiliation{Department of Physics and State Key Laboratory of Nuclear Physics and Technology, Peking University, Beijing, 100871, China}

\author{Chong Sheng Li\footnote{Electronic address: csli@pku.edu.cn}}
\affiliation{Department of Physics and State Key Laboratory of Nuclear Physics and Technology, Peking University, Beijing, 100871, China}
\affiliation{Center for High Energy Physics, Peking University, Beijing, 100871, China}

\author{Hai Tao Li}
\affiliation{Department of Physics and State Key Laboratory of Nuclear Physics and Technology, Peking University, Beijing, 100871, China}



\begin{abstract}
 \vspace*{0.3cm}
ABSTRACT: We investigate threshold resummation effects in the production of a color sextet (antitriplet) scalar at next-to-next-to-leading logarithmic (NNLL) order at the LHC in the frame of soft-collinear effective theory. We show the total cross section and the rapidity distribution with NLO+NNLL accuracy, and we compare them with the NLO results. Besides, we use recent dijet data at the LHC to give the constraints on the couplings between the colored scalars and quarks.
\end{abstract}
\maketitle
\newpage



\section{Introduction}
\label{sec:intro}

There is no clear evidence of new physics beyond the Standard Model found at the LHC so far \cite{Mete:2012cya,VickeyBoeriu:2012nra,Lari:2011ina}, and the most favored supersymmetry, extra dimensions, and many others all receive somewhat strong constraints \cite{Mete:2012cya,Atlas:2012plot,CMS:2012plot}. Then it is a preferable way to be more concerned about the model independent theory rather than considering some specific models. Here we study the model independent color sextet (antitriplet) scalars, which have many significant effects in the phenomenology. Actually, color sextet scalars have been included in many new physics models, such as unification theories \cite{Pati:1974yy,Mohapatra:1980qe,Perez:2007rm}, supersymmetry with R-parity violation \cite{Barbier:2004ez}, and diquark Higgs \cite{Mohapatra:2007af}. Their masses can be as low as the TeV scale or less \cite{Chacko:1998td}, which leads to much impact on the physics. For example, in the supersymmetric Pati-Salam $SU(2)_R\times SU(2)_L\times SU(4)_C$ model, light color sextet scalars can be realized around the weak scale even though the scale of $SU(2)_R\times SU(4)_C$ symmetry breaking is around $10^{10}$ GeV \cite{Chacko:1998td,Mohapatra:2007af}. Observation of the color sextet scalars will be a direct signal of new physics beyond the Standard Model.

Considering the interaction of the color sextet (antitriplet) scalars with quarks, which is parameterized, the relevant Lagrangian can be written as \cite{Han:2009ya}
\begin{equation}\label{eq.Lagrangian}
\mathcal{L}= 2\sqrt{2}\left[\bar{K}_{i}^{ab}\phi^{i}\bar{\psi}_{a}.(\lambda_{L}P_{L}+\lambda_{R}P_{R}).\psi_{b}^{C}+h.c.\right]
+ (D_{\mu}^{ij}\phi_{j})^{\dagger}D_{\mu}^{ik}\phi_{k}-m_{\phi}^{2}\phi^{i\dagger}\phi^{i},
\end{equation}
where $K_i^{ab}$ is the Clebsch-Gordan coefficient of the sextet (antitriplet), $\lambda_{L/R}$ is the Yukawa-like coupling, and $a,b$ are the color indices. The quantum numbers of the colored scalars are listed in Table \ref{table:quantum numbers}, and more information can be found in \cite{Han:2009ya,DelNobile:2009st}.
\begin{table}[h]
  \centering
   \begin{tabular}{c c c c}
     \hline
     \quad $SU(2)_L$ \quad & \quad $U(1)_Y$ \quad & \quad $|Q|=|T_3+Y|$ \quad & \quad couplings to \quad \\
     \hline
     \bf{1} & $1/3$ & $1/3$ & QQ,UD \\
     \bf{3} & $1/3$ & $1/3,2/3,4/3$ & QQ \\
     \bf{1} & $-2/3$ & $2/3$ & DD \\
     \bf{1} & $4/3$ & $4/3$ & UU \\
     \hline
   \end{tabular}
  \caption{Q is the $SU(2)_L$ quark doublet, and U(D) is the up(down)-type $SU(2)_L$ quark singlet. Under $SU(3)_C\times SU(2)_L\times U(1)_Y$, Q has the quantum numbers (\textbf{3},\textbf{2},$1/6$), while U has (\textbf{3},\textbf{1},$2/3$), and D has (\textbf{3},\textbf{1},$-1/3$).}\label{table:quantum numbers}
\end{table}
In order to satisfy the gauge symmetry, the colored scalars couple to same-sign quarks, and then they have fractional electronic charges. In the cases of antitriplet the couplings should be antisymmetric in flavor. For convenience, we label the colored scalars as $sextet^I$, $sextet^{II}$ and $sextet^{III}$ with electronic charge $1/3$, $-2/3$ and $4/3$, respectively. For the antitriplet, the labels are $antitriplet^{I}$, $antitriplet^{II}$ and $antitriplet^{III}$.

It has been shown \cite{Mohapatra:2007af,Chen:2009xjb,Chen:2008hh} that the measurements of $D^0-\bar{D}^0$ mixing and the rate of $D\to\pi^+\pi^0(\pi^+\phi)$ decay can constrain the couplings of the colored scalars to two up-type quarks: $\lambda_R^{uu}, \lambda_R^{uc}\lesssim0.1, |Re(\lambda^{cc}\lambda^{uu*})|\sim 5.76\times 10^{-7}$ for $m_{\phi}\sim $ 1 TeV. Besides, the left-handed coupling $\lambda_L$ also receives a tight constraint due to minimal flavor violation. Since we use the model independent coupling $\lambda^2=\lambda_L^2+\lambda_R^2$, above constraints can be relaxed in the scenario considered below.

Production and decay of the colored scalars at hadron colliders have been extensively discussed in \cite{Chen:2008hh,Han:2009ya,Gogoladze:2010xd,Berger:2010fy,Han:2010rf,Richardson:2011df,Karabacak:2012rn}.
Recently the CMS collaboration has searched for the signal of the colored scalar and obtained limits on the production cross section of such resonant states \cite{CMS:2012yf,Chatrchyan:2013qha,CMS-PAS-EXO-12-059} with the fixed-order theoretical predictions (leading order and next-to-leading order) in Ref. \cite{Hewett:1988xc,Han:2010rf}.
In this paper we investigate the threshold resummation effects in the single production of the color sextet (antitriplet) scalars, and we also discuss the rapidity distribution of the colored scalars at NLO+NNLL accuracy at the LHC with soft-collinear effective theory (SCET) \cite{Bauer:2000yr,Bauer:2001yt,Beneke:2002ph,Bauer:2000ew,Bauer:2002nz}. As a cross check, we also calculate the NLO corrections using the analytical-phase space integral method, and present their analytical expressions. Actually, when the masses of the colored scalars approach the threshold limit, there are large logarithms left after cancelling the divergences, because the scale of the soft gluon radiations is rather small compared to the scalar mass. These threshold logarithms should be resummed to reduce the scale uncertainties and improve the confidence of the theoretical predictions.

This paper is organized as follows: In Sect. \ref{sec:fixed-order calc}, we present the NLO calculations. In Sect. \ref{sec:factorization}, we briefly show the factorization in the threshold limit of the production of the colored scalar. In Sect. \ref{sec:resum}, we calculate the soft function and present solutions of the renormalization group equations obeyed by hard and soft functions. In Sect. \ref{sec:numerical}, we present detailed numerical analyses and compare the NLO+NNLL rapidity distributions with the NLO results. We also use recent dijet data at the LHC to give constraints on the couplings between the colored scalars and quarks. We conclude in Sect. \ref{sec:conclusion}.

\section{Fixed-order calculations}
\label{sec:fixed-order calc}

We consider the process $h_1 +h_2 \rightarrow \phi+X$, where $h_1$ and $h_2$ are the incoming hadrons with momenta $P_1$ and $P_2$, and we define the rapidity of the colored scalar $\phi$ as $Y=\frac{1}{2}\ln\frac{E+p_z}{E-p_z}$, where $E$ and $p_z$ represent the energy and longitudinal momentum of the scalar in the center-of-mass frame of the colliding hadrons. We write the cross section as \cite{Becher:2007ty,Anastasiou:2003ds}:
\begin{equation}\label{eq.differential xsec}
  \frac{d\sigma}{dY} = \sum_{ij} \int_{\tau}^1 \frac{dz}{z} \int_0^1 dy \  f_{i/h_{1}}(x_{1},\mu_f)f_{j/h_{2}}(x_{2},\mu_f) C_{ij}(y,z,m_{\phi},\mu_f),
\end{equation}
\begin{equation}\label{eq.cij000000}
C_{ij}(y,z,m_{\phi},\mu_f) = z\left|\frac{dx_{1}dx_{2}}{dydz}\right| \frac{d\sigma_{ij}}{dY}
  = \frac{1}{2S} \int d\!\ P\!S_{f} \ \overline{|\mathcal{M}_{ij}|^{2}} \ \delta\left(y-\frac{u^{\prime}-z}{(1-z)(1+u^{\prime})}\right),
\end{equation}
with
\begin{eqnarray}
\nonumber
&&  S=(P_1+P_2)^2, \quad \tau=m_{\phi}^2 / S, \quad s=(p_1+p_2)^2=x_1x_2S,\\
&& u^{\prime}=\frac{x_1}{x_2}e^{-2Y}, \quad z=\frac{m_{\phi}^{2}}{s}=\frac{\tau}{x_{1}x_{2}}, \quad y=\frac{u^{\prime}-z}{(1-z)(1+u^{\prime})},
\end{eqnarray}
where $P\!S_f$ is the final state phase space, and $\mu_f$ is the factorization scale. For one-particle final state, there is no $y$ dependence, and then the delta function can be reduced to $(\delta(y)+\delta(1-y))/2$.

The NLO corrections were investigated in Ref. \cite{Han:2009ya} using the phase-space slicing method \cite{Baer:1989xj,Harris:2001sx}. Here we recalculate the matrix elements, which are consistent with the results in Ref. \cite{Han:2009ya}, and we do not present the details of these calculations. Below we just show the analytical expressions of the phase-space integration.
Using the identity
\begin{equation}
x^{-1+\epsilon}=\frac{1}{\epsilon}\delta(x)+\sum_n\frac{\epsilon^n}{n!}\left[\frac{\ln^nx}{x}\right]_+,
\end{equation}
with
\begin{equation}
\int_{\tau}^1dx\left[\frac{\ln^n(1-x)}{1-x}\right]_+f(x)=\int_{\tau}^1dx\frac{\ln^n(1-x)}{1-x}[f(x)-f(1)]-f(1)\int_0^{\tau}dx\frac{\ln^n(1-x)}{1-x},
\end{equation}
we can obtain the following results for $C_{ij}(y,z,m_{\phi},\mu_f)$. The leading-order result is
\begin{equation}
C_{qq}^{(0)} = \frac{2\pi \lambda^2 N_D}{N_C^2 S}\frac{\delta(y)+\delta(1-y)}{2}\delta(1-z),
\end{equation}
and the contributions from the virtual and real corrections for the $qq$ channel are given by
\begin{eqnarray}\label{eq.cqq.virt}
\nonumber
C_{qq}^{virt} &=& \frac{2\pi\lambda^2N_D}{N_C^2S} \frac{\delta(y)+\delta(1-y)}{2}\ \delta(1-z) \frac{\alpha_s}{4\pi}\frac{(4\pi)^{\epsilon}}{\Gamma(1-\epsilon)}
\left[\frac{-4C_F}{\epsilon^2}+\frac{1}{\epsilon}(-2C_D-6C_F+4C_FL) \right.\\
 && \left. +C_D\left(2L-2-\frac{4\pi^2}{3}\right) +C_F\left(-2L^2-4+2\pi^2\right) \right]
\end{eqnarray}
and
\begin{eqnarray}\label{eq.cqq.real}
\nonumber
C_{qq}^{real} &=& \frac{2\pi\lambda^2N_D}{N_C^2S} \frac{\alpha_s}{4\pi}\frac{(4\pi)^{\epsilon}}{\Gamma(1-\epsilon)}
\left\{ \frac{\delta(y)+\delta(1-y)}{2}\delta(1-z) \frac{4C_F}{\epsilon^2} \right.\\
\nonumber
 &&  +\frac{\delta(y)+\delta(1-y)}{2} \frac{1}{\epsilon} \left( (2C_D-4C_FL)\delta(1-z) -8C_F\left[\frac{1}{1-z}\right]_+ +4C_F(1+z) \right) \\
\nonumber
 &&  +C_D\left[\frac{\delta(y)+\delta(1-y)}{2}\delta(1-z)(-2L+4) -4\left[\frac{1}{1-z}\right]_+ +2z+2 \right] \\
\nonumber
 && +C_F\frac{\delta(y)+\delta(1-y)}{2}\left[\delta(1-z)\left(2L^2-\frac{2\pi^2}{3}\right)+8\big(L-\ln(z)\big)\left[\frac{1}{1-z}\right]_+ \right.\\
\nonumber
 &&  \left. +16\left[\frac{\ln(1-z)}{1-z}\right]_+ -4(1+z)\big(L +2\ln(1-z) -\ln(z)\big) +4(1-z) \right] \\
 && \left. +C_F\left(\left[\frac{1}{y}\right]_+ + \left[\frac{1}{1-y}\right]_+\right)\left(4\left[\frac{1}{1-z}\right]_+-2(1+z)\right) \right\},
\end{eqnarray}
respectively.
Combining the contributions of the LO results, the virtual and real corrections, we obtain the bare NLO partonic differential cross sections:
\begin{equation}
C_{qq}^{bare}=C_{qq}^{(0)}+C_{qq}^{virt}+C_{qq}^{real}.
\end{equation}
They still contain the collinear singularities,
which can be factorized into the following form to all orders of perturbation theory in general:
\begin{equation}
C_{ij}^{bare}(z,1/\epsilon)=\sum_{k,l}\Gamma_{ki}(z,\mu_f,1/\epsilon)\otimes\Gamma_{lj}(z,\mu_f,1/\epsilon)\otimes C_{kl}(z,\mu_f),
\end{equation}
where $\mu_f$ is the factorization scale and $\otimes$ is the convolution symbol defined as
\begin{equation}
f(z)\otimes g(z)=\int_z^1\frac{dy}{y}f(y)g(\frac{z}{y}).
\end{equation}
The universal splitting functions $\Gamma_{ij}(z,\mu_f,1/\epsilon)$ represent the probability of finding a particle
$i$ with fraction $z$ of the longitudinal momentum inside the parent particle $j$ at the scale $\mu_f$.
They contain the collinear divergences, and they can be absorbed into the redefinition of the PDF according to mass factorization \cite{Collins:1985ue,Bodwin:1984hc}. Adopting the $\overline{M\!S}$ mass-factorization scheme, we have to $\mathcal{O}(\alpha_s)$
\begin{equation}
\Gamma_{ij}(z,\mu_f,1/\epsilon)=\delta_{ij}\delta(1-z)-\frac{1}{\epsilon}\frac{\alpha_s}{2\pi}\frac{\Gamma(1-\epsilon)}{\Gamma(1-2\epsilon)}\left(\frac{4\pi\mu_r^2}{\mu_f^2}\right)^{\epsilon}
P_{ij}^{(0)}(z),
\end{equation}
where $P_{ij}^{(0)}(z)$ are the leading-order Altarelli-Parisi splitting functions \cite{Altarelli:1977zs}
\begin{eqnarray}
\nonumber
P_{qq}^{(0)}(z)&=&\frac{4}{3}\left[\frac{1+z^2}{(1-z)_+}+\frac{3}{2}\delta(1-z)\right],\\
P_{qg}^{(0)}(z)&=&\frac{1}{2}[(1-z)^2+z^2].
\end{eqnarray}
After absorbing the splitting functions $\Gamma_{ij}(z,\mu_f,1/\epsilon)$ into the redefinition of the PDFs
through the mass factorization in this way, we have the hard-scattering partonic differential cross sections
$C_{ij}(y,z,m_{\phi},\mu_f)$, which are free of collinear divergences, and depend on the scale $\mu_f$.
The final NLO results for the $qq$ channel are given by
\begin{eqnarray}
\nonumber
C_{qq}^{(1)} &=& \frac{2\pi \lambda^2 N_D}{N_C^2 S} \frac{\alpha s}{4\pi} \bigg\{ \frac{\delta(y)+\delta(1-y)}{2}
  \bigg[ \delta(1-z) \bigg( C_D \big(2 - \frac{4}{3} \pi^2 \big) + C_F \big( \frac{4}{3} \pi^2 - 4 \big) \bigg) \\
\nonumber
  && - 8C_F(\ln z-L)\bigg[\frac{1}{1-z}\bigg]_{+} + 16C_F\bigg[\frac{\ln(1-z)}{1-z}\bigg]_{+}
   -4C_F\bigg((z+1)L + 2(z+1)\ln(1-z) \\
\nonumber
  && - (z+1)\ln z + z - 1\bigg) \bigg]
   - 2C_F\bigg[\frac{1}{y}\bigg]_{+}\bigg( -2\bigg[\frac{1}{1-z}\bigg]_{+} +z+1 \bigg) \\
  && - 2C_F\bigg[\frac{1}{1-y}\bigg]_{+}\bigg( -2\bigg[\frac{1}{1-z}\bigg]_{+} +z+1 \bigg)
   + 2C_D\bigg(-2\bigg[\frac{1}{1-z}\bigg]_{+}+z+1\bigg) \bigg\}.
\end{eqnarray}
Similarly, the final NLO result for the $qg$ channel is given by
\begin{eqnarray}
\nonumber
C_{qg}^{(1)} &=& \frac{2\pi\lambda^2 N_D}{N_C(N_C^2-1)S}\frac{\alpha_s}{4\pi}\bigg\{\delta(1-y) 2C_F\bigg[(2z^2-2z+1)\big(L+2\ln{(1-z)}-\ln{z}-1\big)+1\bigg]\\
\nonumber
&& +2C_F\bigg[\frac{1}{1-y}\bigg]_+(2z^2-2z+1) + \frac{2(1-z)}{(yz-y-z)^2} \bigg[C_D\big(y^2(z-1)^2+z^2\big)\\
&& +C_F(y+1)(z-1)(yz-y-z)^2\bigg]\bigg\},
\end{eqnarray}
where $\lambda^2=\lambda_L^2+\lambda_R^2$, $L=\ln(m_{\phi}^2/\mu_{f}^2)$. The color factors are $N_D=6, C_D=10/3$ for the sextet and $N_D=3, C_D=4/3$ for the antitriplet. In the above results, we have set the renormalization scale $\mu_r=\mu_f$.
Finally, we combine these finite results to arrive at the NLO differential cross section $C_{ij}(y,z,m_{\phi},\mu_f)$ for colored scalar production:
\begin{equation}
C_{ij}=C_{qq}^{(0)}+C_{qq}^{(1)}+C_{qg}^{(1)}.
\end{equation}

Following the method in \cite{Becher:2007ty}, we rearrange the results as
\begin{equation}\label{eq.integral kernel}
C_{qq}(z,y,m_{\phi},\mu_f)=C_{qq}^{(0)}+C_{qq}^{(1)}=\frac{2\pi N_D}{S N_C^2}\frac{\delta(y)+\delta(1-y)}{2}C(z,m_{\phi},\mu_f)+C_{qq}^{\text{subleading}},
\end{equation}
where the $C(z,m_{\phi},\mu_f)$ are the leading singular terms (threshold terms), which are arranged as
\begin{eqnarray}\label{eq.fixed-order thres}
\nonumber
  C(z,m_{\phi},\mu_f) &=& \lambda^2\delta[1-z] + \lambda^2\frac{\alpha_s}{4\pi}\bigg\{
  \delta[1-z]\left[C_D\left(2-\frac{4}{3}\pi^2\right)+C_F\left(-4+\frac{4}{3}\pi^2\right)\right] \\
  && + \left[\frac{1}{1-z}\right]_{+}\big[-4C_D+8C_F(L-\ln z)\big]
     +  \left[\frac{\ln (1-z)}{1-z}\right]_{+}16C_F \bigg\}.
\end{eqnarray}
From Eq. (\ref{eq.fixed-order thres}), we can see that the singular terms make the perturbative series badly convergent in the
threshold limit $z\rightarrow 1$, and thus they must be resummed to all orders.

\section{Factorization at threshold in SCET}
\label{sec:factorization}

The production of the colored scalar involves several scales, which are
\begin{equation}
s,m_{\phi}^2\gg s(1-z)^2\gg\Lambda_{QCD}^2
\end{equation}
in the threshold limit, and  it is convenient to introduce two light-like vectors $n$ and $\bar{n}$ along the directions of the colliding partons, which satisfy $n\cdot\bar{n}=2$.
In the lab frame, they can be written as
\begin{equation}
n=(1,0,0,1),\quad\bar{n}=(1,0,0,-1).
\end{equation}
Then any four vector can be decomposed as
\begin{equation}
k^{\mu}=n\cdot k\frac{\bar{n}^{\mu}}{2}+\bar{n}\cdot k\frac{n^{\mu}}{2}+k_{\perp}^{\mu}\equiv k^+\frac{\bar{n}^{\mu}}{2}+k^-\frac{n^{\mu}}{2}+k_{\perp}^{\mu}.
\end{equation}
In this limit, we need to distinguish four different momentum regions
\begin{eqnarray}
\nonumber
\text{hard:} && k^{\mu}\sim\sqrt{s}(1,1,1),\\
\nonumber
\text{hard-collinear:} && k^{\mu}\sim\sqrt{s}(\epsilon,1,\sqrt{\epsilon}),\\
\nonumber
\text{anti-hard-collinear:} && k^{\mu}\sim\sqrt{s}(1,\epsilon,\sqrt{\epsilon}),\\
\text{soft:} && k^{\mu}\sim\sqrt{s}(\epsilon,\epsilon,\epsilon),
\end{eqnarray}
where we use $k^{\mu}=(k^+,k^-,k_{\perp})$ to denote the momenta and $\epsilon=(1-z)\ll 1$.
Generally, the differential cross section can be written as
\begin{equation}
d\sigma=\frac{1}{2S}\frac{d^3\vec{q}}{(2\pi)^32E_{\phi}}\int d^4x \langle N_1(P_1)N_2(P_2)|\mathcal{H}_{eff}^{\dagger}(x)|\phi(q)\rangle
\langle\phi(q)|\mathcal{H}_{eff}(0)|N_1(P_1)N_2(P_2)\rangle,
\end{equation}
where the effective Hamiltonian is given by
\begin{equation}
\mathcal{H}_{eff}(x)=\int dt_1 dt_2\ e^{im_{\phi}v\cdot x}\ \tilde{C}(t_1,t_2)\ \mathcal{O}(x,t_1,t_2),
\end{equation}
with
\begin{equation}
\mathcal{O}(x,t_1,t_2)=2\sqrt{2}\ Y_{\bar{n}}^{a\dagger}\bar{\chi}_{\bar{n}}(x+t_2n).(\lambda_LP_L+\lambda_RP_R).Y_n^{b\dagger}\chi_n^C(x+t_1\bar{n})
\ Y_{v}^i\phi_{v}(x)\ K_i^{ab},
\end{equation}
where $\chi_n$ is the gauge-invariant combination of the $n$-collinear quark field and $n$-collinear Wilson line, and $Y$ is defined as the soft Wilson line \cite{Bauer:2001yt,Hill:2002vw,Becher:2003qh}:
\begin{eqnarray}
\nonumber
Y_n(x) &=& \mathcal{P}\exp\left(ig_s\int_{-\infty}^0 dt_0\ n\cdot A_s^a(x+t_0n)t^a\right),\\
Y_{v}(x) &=& \mathcal{P}\exp\left(-ig_s\int_0^{\infty} dt_0\ v\cdot A_s^a(x+t_0v)t^a\right),
\end{eqnarray}
where $v$ is the velocity of the colored scalar.
The matrix element can be factorized as follows:
\begin{eqnarray}
\nonumber
\langle N_1(P_1)N_2(P_2)|\mathcal{O}^{\dagger}(x)\mathcal{O}(0)|N_1(P_1)N_2(P_2)\rangle
 &=& \frac{2\lambda^2N_D}{N_C^2} \langle N_1(P_1)|\bar{\chi}_n(x)\frac{/\!\!\!\bar{n}}{2}\chi_n(0)|N_1(P_1)\rangle \\
 && \times \langle N_2(P_2)|\bar{\chi}_{\bar{n}}(x)\frac{/\!\!\! n}{2}\chi_{\bar{n}}(0)|N_2(P_2)\rangle\ \hat{\mathcal{W}}(x,\mu_f),
\end{eqnarray}
with
\begin{equation}
\hat{\mathcal{W}}(x,\mu_f)=\frac{1}{N_D}\left\langle 0\left|\ \text{Tr}\ \left(\bar{T}\left[Y_n^{\dagger}(x)Y_{\bar{n}}^{\dagger}(x)Y_{v}(x)\right]\ T\left[Y_{\bar{n}}(0)Y_n(0)Y_{v}^{\dagger}(0)\right]\right)\right|0\right\rangle,
\end{equation}
where the trace is over color indices, and $\bar{T}$ denotes the anti-time-ordering operator.
The initial state collinear sector reduces to the conventional PDFs \cite{Collins:1981uw,Bauer:2001yt}:
\begin{equation}
f_{i/N}(x,\mu)=\frac{1}{2\pi}\int dt\ e^{-ixt\bar{n}\cdot p}\langle N(p)|\bar{\chi}(t\bar{n})\frac{/\!\!\!\bar{n}}{2}\chi(0)|N(p)\rangle.
\end{equation}
The integrals over $t_1$ and $t_2$ produce the Fourier-transformed Wilson coefficients:
\begin{equation}
C_H(-\bar{n}\cdot p_1\ n\cdot p_2,\mu_f)=\int dt_1 dt_2 e^{-it_1\bar{n}\cdot p_1-it_2n\cdot p_2}\tilde{C}(t_1,t_2,\mu_f).
\end{equation}
Finally, the singular differential cross section in the threshold region can be written as
\begin{equation}
\nonumber
\frac{d\sigma}{dY} = \frac{2\pi N_D}{SN_C^2}\ \sum_{i,j}\int_{\tau}^1\frac{dz}{z}\int_0^1 dy\ f_{i/h_1}(x_1,\mu_f)f_{j/h_2}(x_2,\mu_f)  \frac{\delta(y)+\delta(1-y)}{2}\ C(z,m_{\phi},\mu_f).
\end{equation}
Following the approach in ref. \cite{Ahrens:2008nc},
$C(z,m_{\phi},\mu_f)$ can be factorized as
\begin{equation}
C(z,m_{\phi},\mu_f)=\lambda^2(\mu_f)\mathbf{H}(m_{\phi},\mu_f)\mathbf{S}\big(\sqrt{s}(1-z),\mu_f\big),
\end{equation}
with
\begin{eqnarray}
\nonumber
\mathbf{H}(m_{\phi},\mu_f) &=& \big|C_H(-m_{\phi}^2-i\epsilon,\mu_f)\big|^2, \\
\nonumber
\mathbf{S}(\sqrt{s}(1-z),\mu_f) &=& \sqrt{s}\ \mathcal{W}(\sqrt{s}(1-z),\mu_f), \\
\mathcal{W}(\omega,\mu_f) &=& \int\frac{dx^0}{4\pi}\ e^{i\omega x^0/2}\ \hat{\mathcal{W}}(x^0,\vec{x}=0,\mu_f).
\end{eqnarray}
The soft and collinear degrees of freedom decouple in the threshold limit, so the physics at different scales can be studied separately \cite{Bauer:2001yt}.

\section{Resummation}
\label{sec:resum}

The coupling $\lambda$ satisfies the renormalization group equation
\begin{equation}
\frac{d}{d\ln\mu}\lambda(\mu)=\gamma^{\lambda}(\alpha_s)\lambda(\mu),
\end{equation}
where the one-loop level $\gamma^{\lambda}$ is
\begin{equation}
\gamma^{\lambda}=-\frac{\alpha_s}{4\pi}6C_F.
\end{equation}
The hard function encodes short distance information
\begin{equation}\label{eq.C_H.perturbative.expansion}
\mathbf{H}(m_{\phi},\mu_f)=\big|C_H(-m_{\phi}^2-i\epsilon,\mu_f)\big|^2
=1+\sum_{n=1}^{\infty}c_n(L)\left(\frac{\alpha_s(\mu_f)}{4\pi}\right)^n.
\end{equation}
We read off the results from the virtual correction:
\begin{equation}
\mathbf{H}(m_{\phi},\mu_f)=1+\frac{\alpha_s}{4\pi}\left[C_D\left(2L-\frac{4}{3}\pi^2-2\right)+
C_F\left(-2L^2+\frac{7}{3}\pi^2-4\right)\right].
\end{equation}
$C_H$ satisfies the RGE \cite{Becher:2007ty}
\begin{equation}\label{eq.C_V.RGE}
\frac{d}{d\ln\mu}C_H(-m^2-i\epsilon,\mu) =\left[\Gamma_{\text{cusp}}(\alpha_s)\left(\ln\frac{m^2}{\mu^2}-i\pi\right)+\gamma^H(\alpha_s)\right]C_H(-m^2-i\epsilon,\mu),
\end{equation}
with
\begin{equation}
\gamma^H=2\gamma^q+\gamma^D-\gamma^{\lambda}.
\end{equation}
$\gamma^q$ is the anomalous dimension of the massless quark \cite{Gehrmann:2005pd}, and $\gamma^D$ is the one of the final state colored scalar, which is given by \cite{Becher:2009kw}
\begin{eqnarray}
\nonumber
&&  \gamma_0^D = -2C_D, \\
&&  \gamma_1^D = C_DC_A\left(\frac{2\pi^2}{3}-\frac{98}{9}-4\zeta_3\right) +\frac{40}{9}C_DT_Fn_f.
\end{eqnarray}
The solution of Eq.(\ref{eq.C_V.RGE}) is \cite{Becher:2007ty}
\begin{equation}
C_H(-m_{\phi}^2,\mu_f)=\exp\left[2S(\mu_h,\mu_f)-a_{\gamma^{H}}(\mu_h,\mu_f)+i\pi a_{\Gamma}(\mu_h,\mu_f)\right]\left(\frac{m_{\phi}^2}{\mu_h^2}\right)^{-a_{\Gamma}(\mu_h,\mu_f)}C_H(-m_{\phi}^2,\mu_h),
\end{equation}
with
\begin{equation}\label{eq.Sudakov.etc.S}
S(\nu,\mu)=-\int_{\alpha_s(\nu)}^{\alpha_s(\mu)}d\alpha\frac{\Gamma_{\text{cusp}}(\alpha)}{\beta(\alpha)}
\int_{\alpha_s(\nu)}^{\alpha}\frac{d\alpha^{\prime}}{\beta(\alpha^{\prime})},
\end{equation}
\begin{equation}\label{eq.Sudakov.etc.aG}
a_{\Gamma}(\nu,\mu)=-\int_{\alpha_s(\nu)}^{\alpha_s(\mu)}d\alpha\frac{\Gamma_{\text{cusp}}(\alpha)}{\beta(\alpha)},
\end{equation}
where $\mu_h $ is the hard matching scale, and for $a_{\gamma^{H}}$ we have a similar expression.

\begin{figure}[h]
  \includegraphics[scale=1.2]{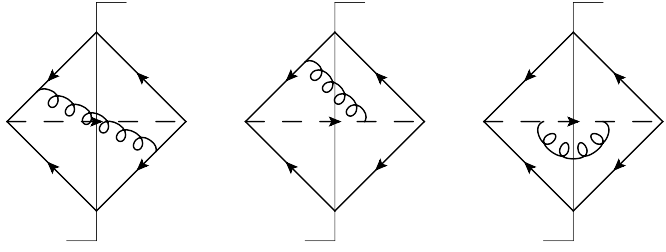}
  \caption{Diagrams for calculating the $\mathcal{O}(\alpha_s)$ soft function. The solid lines represent Wilson lines in the light-like $n$ and $\bar{n}$ directions, the dashed lines represent Wilson lines in the $v$ direction, and the cut curly lines represent the cut gluon propagators.}\label{figsofteikonal}
\end{figure}
Up to the NLO, the soft matrix elements accounting for soft gluon radiations from initial and final states can be obtained after calculating the Feynman diagrams shown in Fig. \ref{figsofteikonal}, and the soft function is given by
\begin{eqnarray}
\nonumber
\mathbf{S}(\sqrt{s}(1-z),\mu_f) &=& \frac{\alpha_s}{4\pi}\biggl[\delta(1-z)\bigl(C_D(-2L+4)+C_F(2L^2-\pi^2)\bigr) \\
 && +\big(-4C_D+8C_F(L-\ln z)\big)\left[\frac{1}{1-z}\right]_{+}+16C_F\left[\frac{\ln(1-z)}{1-z}\right]_{+}\biggr].
\end{eqnarray}
It satisfies the RGE \cite{Becher:2007ty}
\begin{eqnarray}\label{eq.W.RGE}
\nonumber
\frac{d\ \mathcal{W}(\omega,\mu)}{d\ln\mu} &=& -\left[4\Gamma_{\text{cusp}}(\alpha_s)\ln\frac{\omega}{\mu}+2\gamma^W(\alpha_s)\right]
\mathcal{W}(\omega,\mu) \\
 && -4\Gamma_{\text{cusp}}(\alpha_s)\int_{0}^{\omega}d\omega^{\prime}\frac{\mathcal{W}(\omega^{\prime},\mu) -\mathcal{W}(\omega,\mu)}{\omega-\omega^{\prime}},
\end{eqnarray}
with
\begin{equation}
\gamma^W=2\gamma^{\phi}+\gamma^H+\gamma^{\lambda},
\end{equation}
where $\gamma^{\phi}$ is the anomalous dimension of the PDF \cite{Moch:2004pa}. Its solution is \cite{Becher:2007ty}
\begin{equation}\label{eq.soft.RGE.solution}
\mathcal{W}(\omega,\mu_f) =exp\bigl[-4S(\mu_s,\mu_f)+2a_{\gamma^W}(\mu_s,\mu_f)\bigr]\tilde{s}(\partial_{\eta},\mu_s)\frac{1}{\omega}
\left(\frac{\omega}{\mu_s}\right)^{2\eta}\frac{e^{-2\gamma_E \eta}}{\Gamma(2\eta)},
\end{equation}
with
\begin{equation}
\eta=2a_{\Gamma}(\mu_s,\mu_f),
\end{equation}
where $\partial_{\eta}$ is the derivative with respect to $\eta$, and $\tilde{s}$ is obtained by a Laplace transformation
\begin{equation}
\tilde{s}(L,\mu_s)=\int_{0}^{\infty}d\omega e^{-s\omega}\mathcal{W}(\omega,\mu_s)=1+\frac{\alpha_s}{4\pi}\biggl[C_D(-2L+4)+C_F\bigl(2L^2+\frac{4}{3}\pi^2\bigr)\biggr],
\end{equation}
with
\begin{equation}
s=\frac{1}{e^{\gamma_E}\mu_s e^{L/2}}.
\end{equation}

Combining the above formulae, the RG-improved integral kernel is given by
\begin{eqnarray}\label{eq_resummed_C}
\nonumber
  C(z,m_{\phi},\mu_f) &=& \lambda^2(\mu_{\lambda})|C_H(-m_{\phi}^2,\mu_h)|^2 U(m_{\phi},\mu_{\lambda},\mu_h,\mu_s,\mu_f) \\
  && \cdot\frac{z^{-\eta}}{(1-z)^{1-2\eta}}\tilde{s}\left(\ln\frac{m_{\phi}^2(1-z)^2}{\mu_s^2 z}+\partial_{\eta}\ ,\ \mu_s\right)  \frac{e^{-2\gamma_E \eta}}{\Gamma(2\eta)},
\end{eqnarray}
with
\begin{eqnarray}\label{eq.evolution.U}
\nonumber
U(m,\mu_{\lambda},\mu_h,\mu_s,\mu_f) &=& \left(\frac{m^2}{\mu_h^2}\right)^{-2a_{\Gamma}(\mu_h,\mu_s)}\\
&& \times \exp\left[4S(\mu_h,\mu_s)+4a_{\gamma^{\phi}}(\mu_s,\mu_f)-2a_{\gamma^{H}}(\mu_h,\mu_s)
-2a_{\gamma^{\lambda}}(\mu_{\lambda},\mu_s)\right].
\end{eqnarray}

\begin{table}[h]
  \centering
   \begin{tabular}{c c c c c c}
     \hline
     \quad RG-impr.PT \quad & \quad Log.approx \quad &\quad  Accuracy$\sim\alpha_s^nL^k$ \quad &\qquad  $\Gamma_{\text{cusp}}$ \qquad & \qquad $\gamma^H,\gamma^{\phi},\gamma^{\lambda}$ \qquad & \qquad $C_H,\tilde{s}$ \qquad \\
     \hline
     - & LL & $k=2n$ & \qquad 1-loop & \qquad tree-level & \qquad tree-level \\
     LO & NLL & $2n-1\leq k\leq 2n$ & \qquad 2-loop & \qquad 1-loop & \qquad tree-level \\
     NLO & NNLL & $2n-3\leq k\leq 2n$ & \qquad 3-loop & \qquad 2-loop & \qquad 1-loop \\
     \hline
   \end{tabular}
  \caption{Schemes for resummation with different levels of accuracy.}
  \label{Table.approx.schemes}
\end{table}
For convenience, we list the counting scheme in Table \ref{Table.approx.schemes}, which shows corresponding requirements of different levels of accuracy \cite{Becher:2007ty}.
Currently the two-loop $\gamma^{\lambda}$ is not available in the literature, so we just use the one-loop $\gamma^{\lambda}$. The contribution of $\gamma^{\lambda}$ in the evolution function $U(m,\mu_{\lambda},\mu_h,\mu_s,\mu_f)$ cancels out when $\mu_{\lambda}\sim\mu_h$, so $\gamma^{\lambda}$ only affects the running of $\lambda(\mu_{\lambda})$, which gives a subordinate contribution. We then call our resummation an approximate next-to-next-to-leading logarithmic ($\text{NNLL}_{\text{approx}}$), which is combined with the NLO results as follows:
\begin{equation}\label{eq.combined resummation}
\frac{d\sigma^{\text{combined}}}{dY}=\frac{d\sigma^{\text{thresh}}}{dY}\bigg|_{\mu_{\lambda},\mu_h,\mu_s,\mu_f}
+\left(\frac{d\sigma^{\text{fixed-order}}}{dY}\bigg|_{\mu_f}
-\frac{d\sigma^{\text{thresh}}}{dY}\bigg|_{\mu_{\lambda}=\mu_h=\mu_s=\mu_f}\right).
\end{equation}

\section{Numerical Discussion}
\label{sec:numerical}

In this section, we discuss the numerical results for threshold resummation effects in the single production of the color sextet (antitriplet) scalars at the LHC. Throughout our work the PDFsets MSTW2008lo and MSTW2008nlo \cite{Martin:2009iq,Martin:2009bu,Martin:2010db} are used for LO, NLL and NLO, NNLL$_{\text{approx}}$, respectively. If not explained specially, we will assume the coupling $\lambda^2(M_Z)=0.01\alpha_s(M_Z)$, and we choose the initial state quarks $uu$ for the sextet and $ud$ for the antitriplet.

\begin{figure}[h]
  \centering
  \includegraphics[scale=0.8]{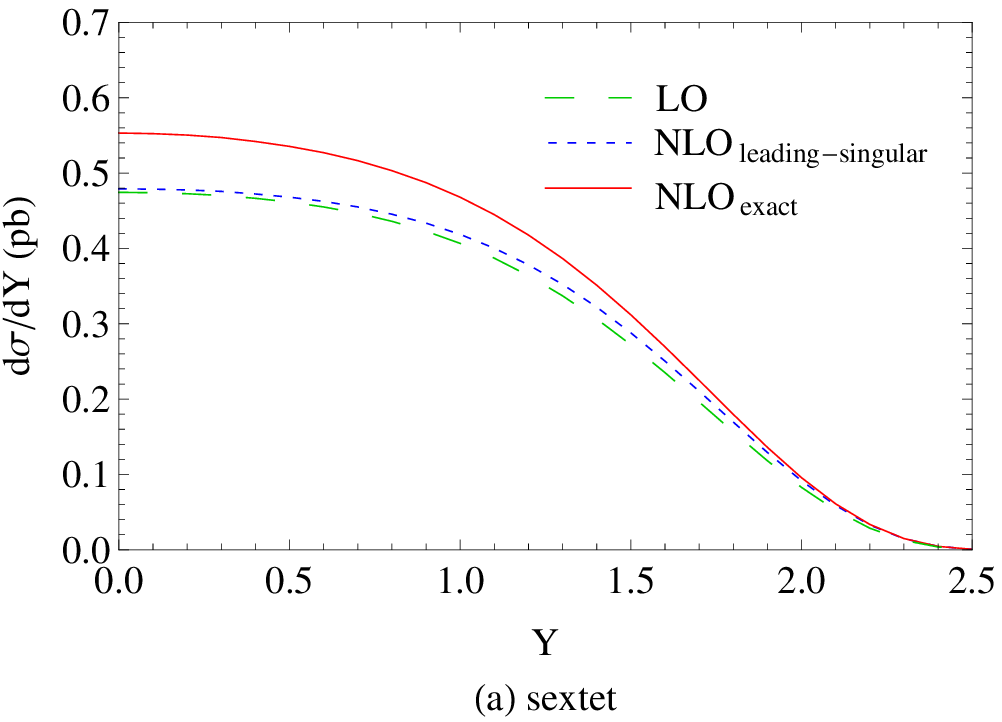}
  \includegraphics[scale=0.8]{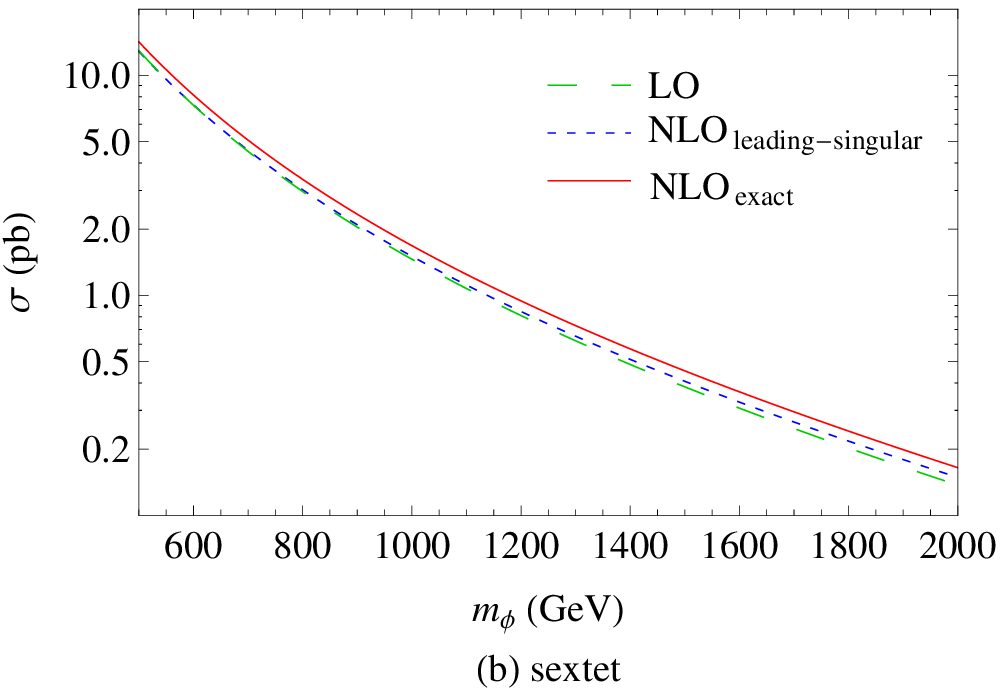}\\
  \includegraphics[scale=0.8]{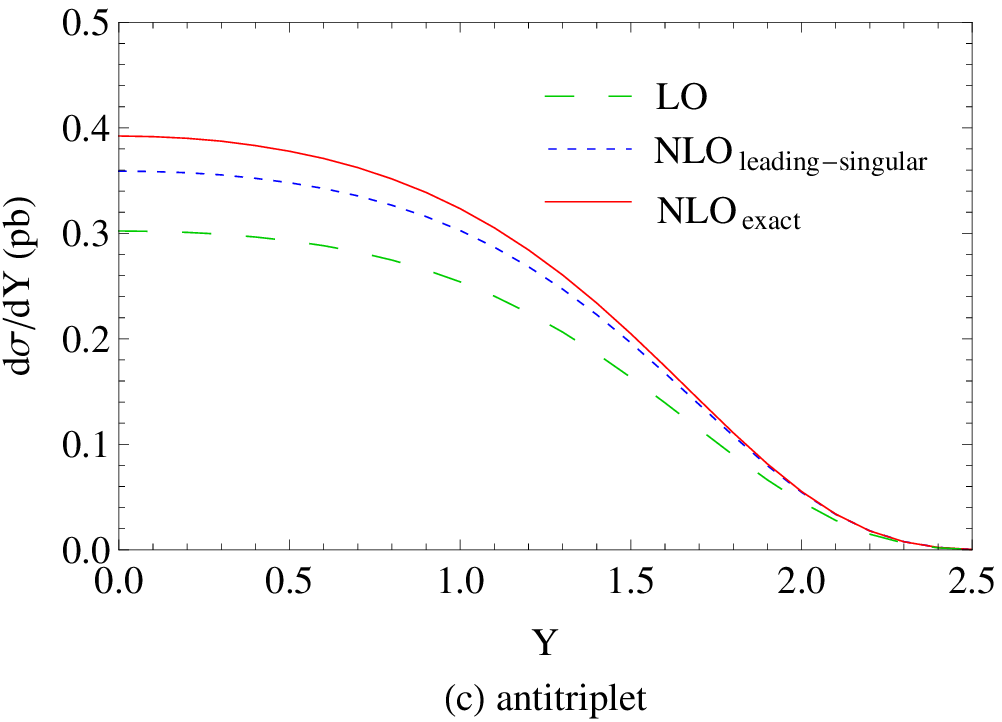}
  \includegraphics[scale=0.8]{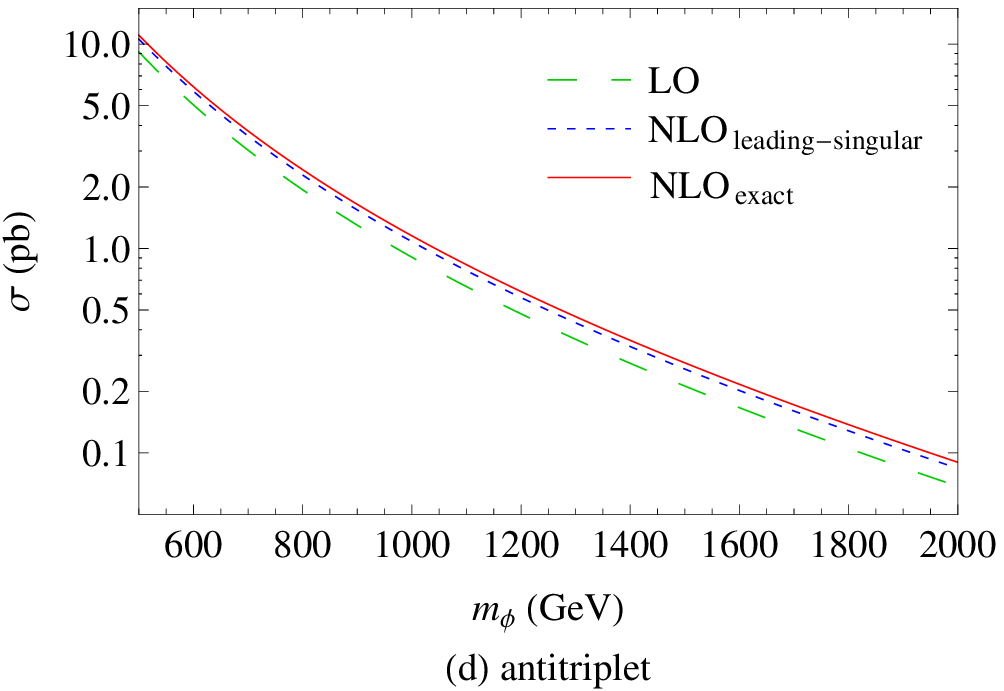}\\
  \caption{Comparison of the exact NLO results and the leading singular results. The long-dashed, dashed and solid lines correspond to LO, leading singular NLO and exact NLO results, respectively. The mass of the colored scalars is set to be $1$\ TeV in the rapidity distributions, and the center-of-mass energy of the colliding hadrons is set to be $14$ TeV.}\label{fig.compare.singular}
\end{figure}
The comparison between the leading singular results and the NLO results is shown in Fig. \ref{fig.compare.singular}. We find that the leading singular terms give the dominant contribution, and the leading singular contribution of the sextet is smaller than the one of the antitriplet. The reason is that the terms associated with $C_D$ give a negative contribution, and then a larger $C_D$ of the sextet leads to smaller leading singular results.

\begin{figure}[h]
  \centering
  \includegraphics[scale=0.8]{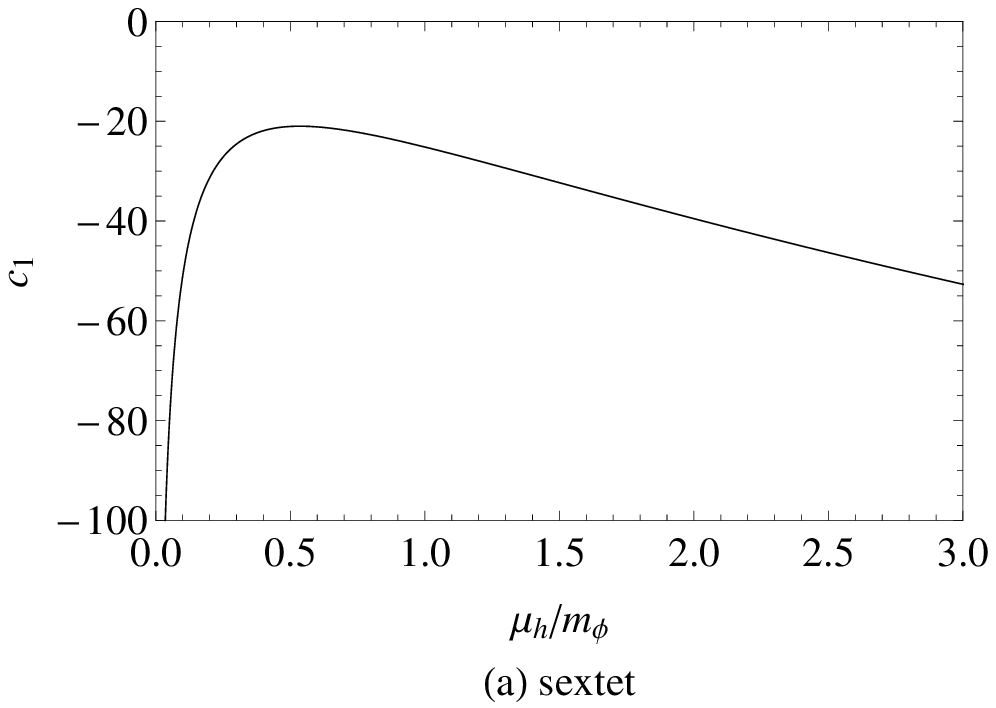}
  \includegraphics[scale=0.8]{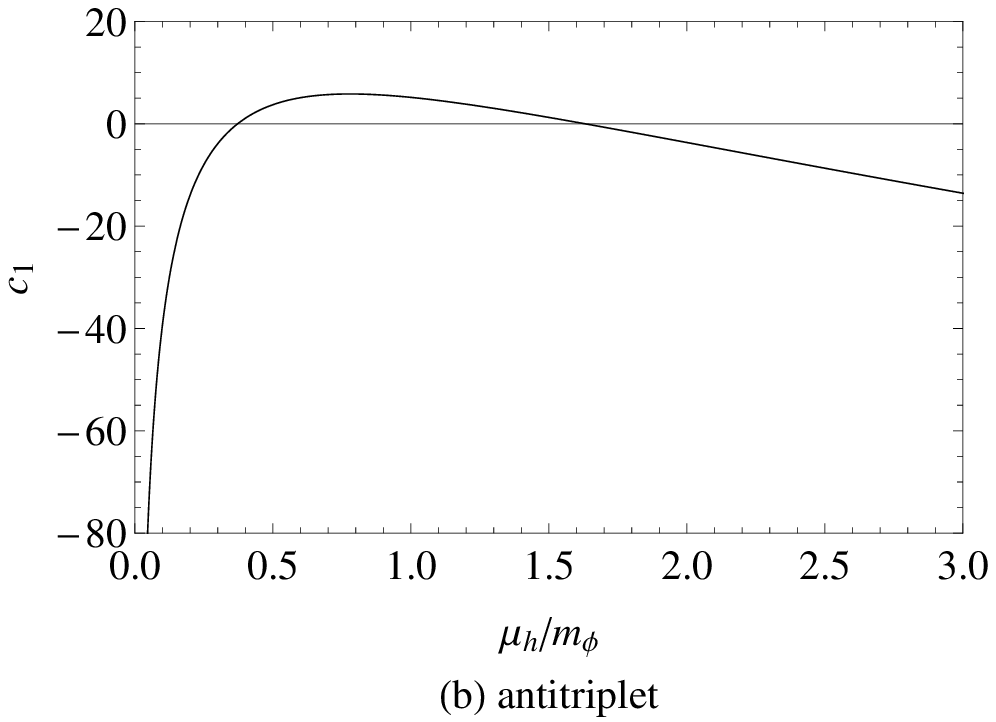}
  \caption{The $\mu_h$ dependence of the expansion coefficients $c_1$ in the hard function.}\label{fig.find mu h}
\end{figure}

Taking the perturbative convergence of $C_H$ and $\tilde{s}$ as the guiding principle, we can obtain the matching scales $\mu_h$ and $\mu_s$. In Fig. \ref{fig.find mu h} we show the $\mu_h$ dependence of the expansion coefficient $c_1$ defined in Eq.(\ref{eq.C_H.perturbative.expansion}). We choose the hard scale $\mu_h^0=0.535m_{\phi}$ for the sextet
and $\mu_h^0=1.63m_{\phi}$ for the antitriplet, respectively.
\begin{figure}[h]
  \includegraphics[scale=0.8]{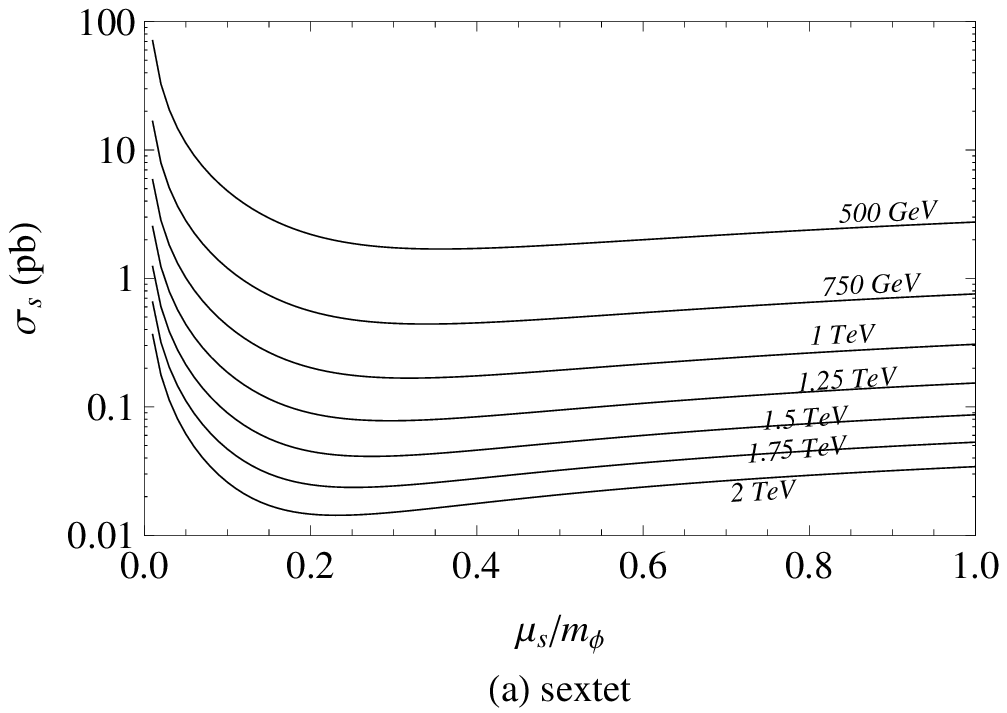}
  \includegraphics[scale=0.8]{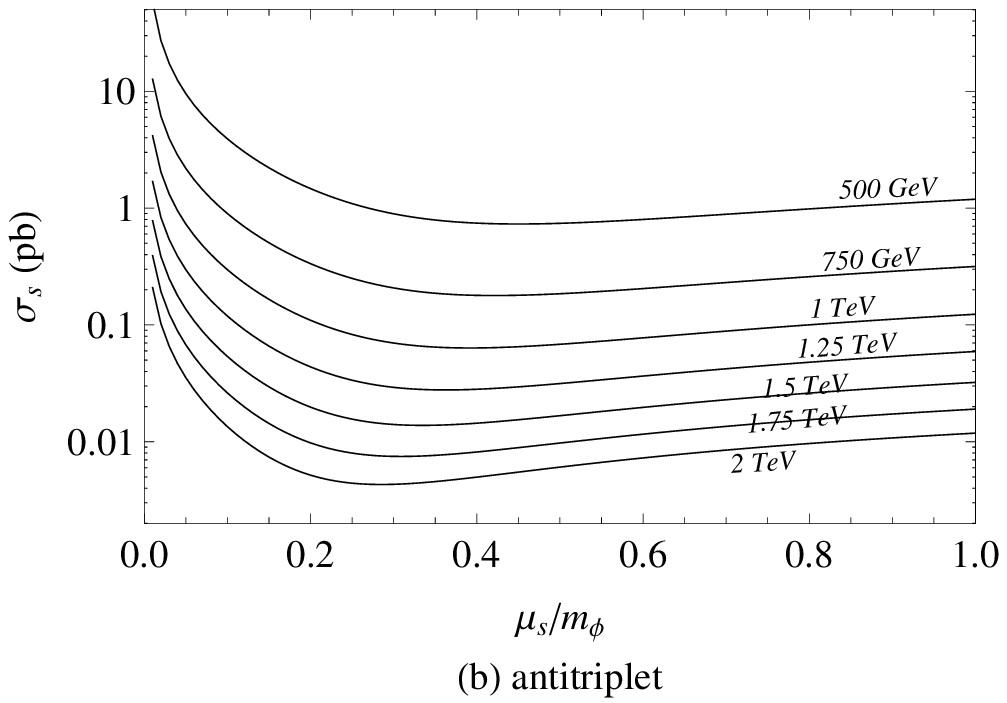}
  \caption{The $\mu_s$ dependence of the soft function with different masses of the colored scalars.}\label{fig.findmus}
\end{figure}
The $\mu_s$ dependence of the soft function is shown in Fig. \ref{fig.findmus}. We fit the results and obtain the empirical functions:
\begin{eqnarray}\label{}
\nonumber
\text{sextet:} && \mu_s^0=\frac{m_{\phi}(1-\tau)}{\sqrt{7+540\tau}}, \\
\text{antitriplet:} && \mu_s^0=\frac{m_{\phi}(1-\tau)}{\sqrt{4.6+362\tau}}.
\end{eqnarray}
It is required that $\mu_{\lambda}$ reflects the intensity of the interaction between the colored scalars and quarks, and $\mu_{\lambda}=\mu_h$ is reasonable.

\begin{table}[h]
  \centering
   \begin{tabular}{c c c c c c c}
     \hline\hline
      \multicolumn{7}{c}{sextet} \\
      \hline
       & \multicolumn{3}{c}{$\sqrt{S}$=8 TeV} & \multicolumn{3}{c}{$\sqrt{S}$=14 TeV} \\
      \quad $m_{\phi}$\quad  & \quad LO\quad & \quad NLO\quad  & \quad NLO+NNLL$_{\text{approx}}$\quad  & \quad LO\quad & \quad NLO\quad  & \quad NLO+NNLL$_{\text{approx}}$\quad  \\
      \quad 0.5 TeV\quad   & \quad 7.53\quad & \quad 8.59\quad  & \quad 8.58\quad  & \quad 12.9\quad & \quad 14.2\quad  & \quad 14.2\quad  \\
      \quad 1 TeV\quad   & \quad 0.768\quad & \quad 0.916\quad  & \quad 0.918\quad  & \quad 1.46\quad & \quad 1.68\quad  & \quad 1.68\quad  \\
      \quad 2 TeV\quad   & \quad 0.0416\quad & \quad 0.0512\quad  & \quad 0.0529\quad  & \quad 0.137\quad & \quad 0.165\quad  & \quad 0.165\quad  \\
     \hline
      \multicolumn{7}{c}{antitriplet} \\
      \hline
       & \multicolumn{3}{c}{$\sqrt{S}$=8 TeV} & \multicolumn{3}{c}{$\sqrt{S}$=14 TeV} \\
      \quad $m_{\phi}$\quad  & \quad LO\quad & \quad NLO\quad  & \quad NLO+NNLL$_{\text{approx}}$\quad  & \quad LO\quad & \quad NLO\quad  & \quad NLO+NNLL$_{\text{approx}}$\quad  \\
      \quad 0.5 TeV\quad   & \quad 4.85\quad & \quad 6.13\quad  & \quad 6.21\quad  & \quad 9.17\quad & \quad 11.1\quad  & \quad 11.2\quad  \\
      \quad 1 TeV\quad   & \quad 0.406\quad & \quad 0.532\quad  & \quad 0.542\quad  & \quad 0.907\quad & \quad 1.15\quad  & \quad 1.17\quad  \\
      \quad 2 TeV\quad   & \quad 0.0161\quad & \quad 0.0215\quad  & \quad 0.0225\quad  & \quad 0.0686\quad & \quad 0.899\quad  & \quad 0.916\quad  \\
     \hline\hline
   \end{tabular}
  \caption{Numerical results of the total cross section (unit: pb). }\label{table.tot.xsec}
\end{table}

In Table \ref{table.tot.xsec}, we list the typical results of total cross sections, which compare NLO+NNLL$_{\text{approx}}$ with LO and NLO results. From Table \ref{table.tot.xsec}, we can see that the resummation effects increase the NLO total cross section by about $2\%$ and $0.2\%$ for 1 TeV antitriplet and sextet, respectively, and $5\%$ and $3\%$ for 2 TeV antitriplet and sextet, respectively, at the 8 TeV LHC.
And the resummation effects at the 14 TeV LHC are smaller than the ones at the 8 TeV LHC.

\begin{figure}[h]
  \centering
  \includegraphics[scale=0.5]{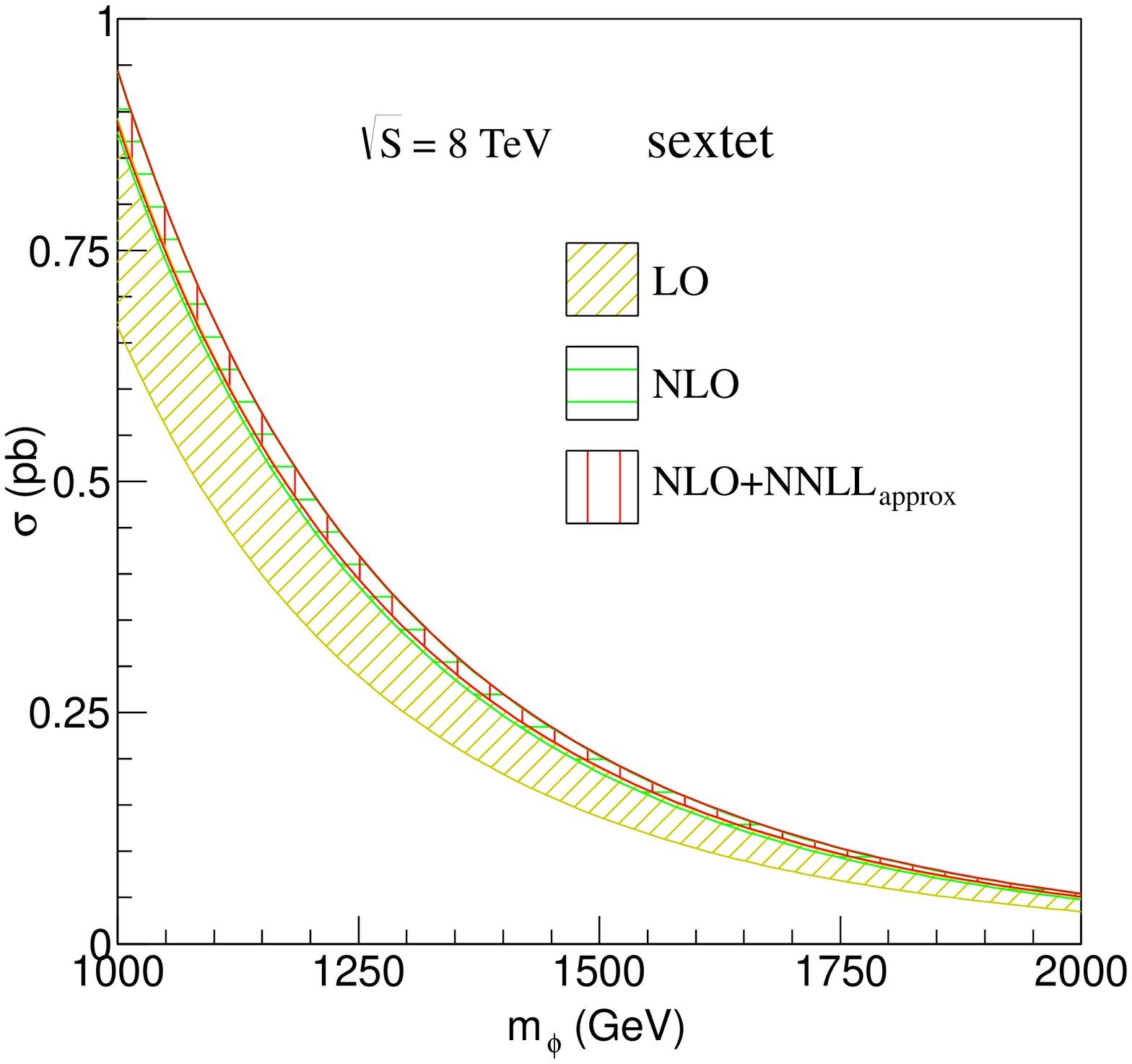}\\
  \includegraphics[scale=0.5]{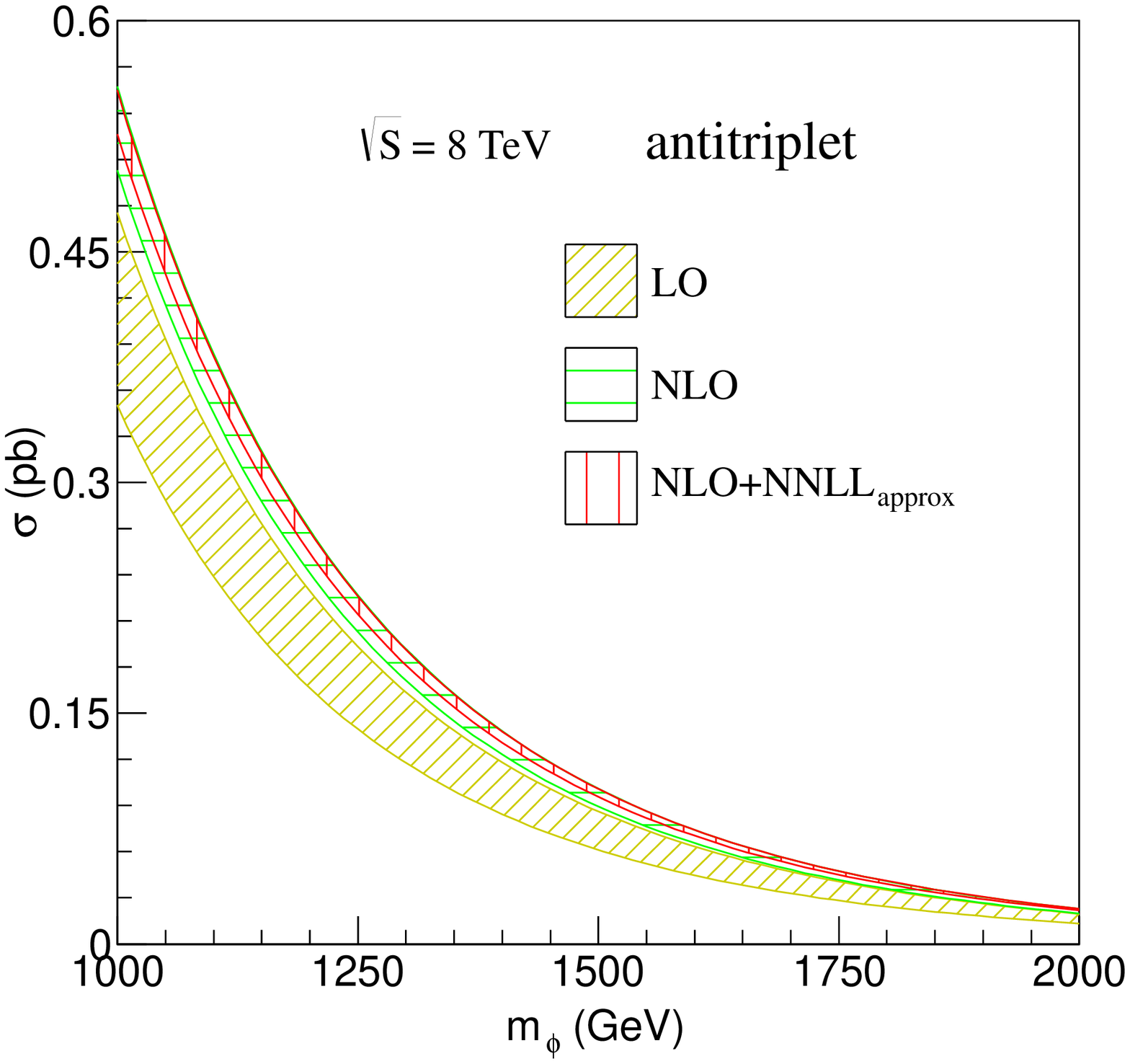}\\
  \caption{The fixed-order and RG-improved cross section predictions including perturbative uncertainty bands due to variations of scale $\mu_f$.}\label{fig.m.trend}
\end{figure}

In Fig. \ref{fig.m.trend}, we show the dependence of the total cross section on the scalar masses including perturbative uncertainty bands due to variation of scale $\mu_f$ at the 8 TeV LHC. We find that the threshold resummation reduces the scale dependence of the total cross section. The scenario at the 14 TeV LHC is very similar, so we do not present it in the figures.

\begin{figure}[h]
  \centering
  \includegraphics[scale=0.8]{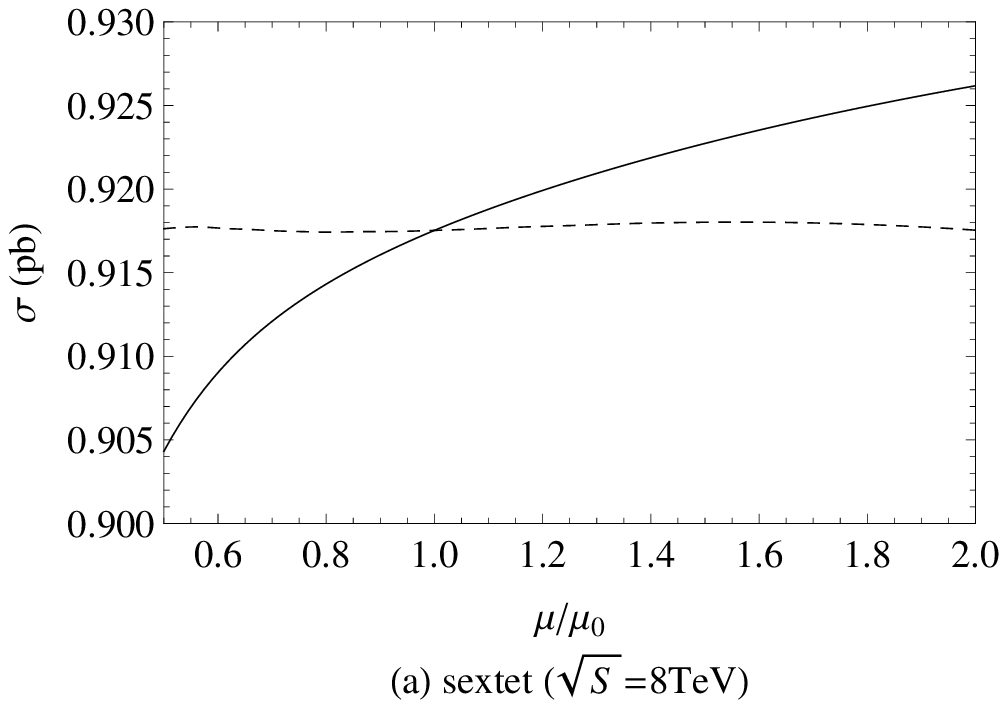}
  \includegraphics[scale=0.8]{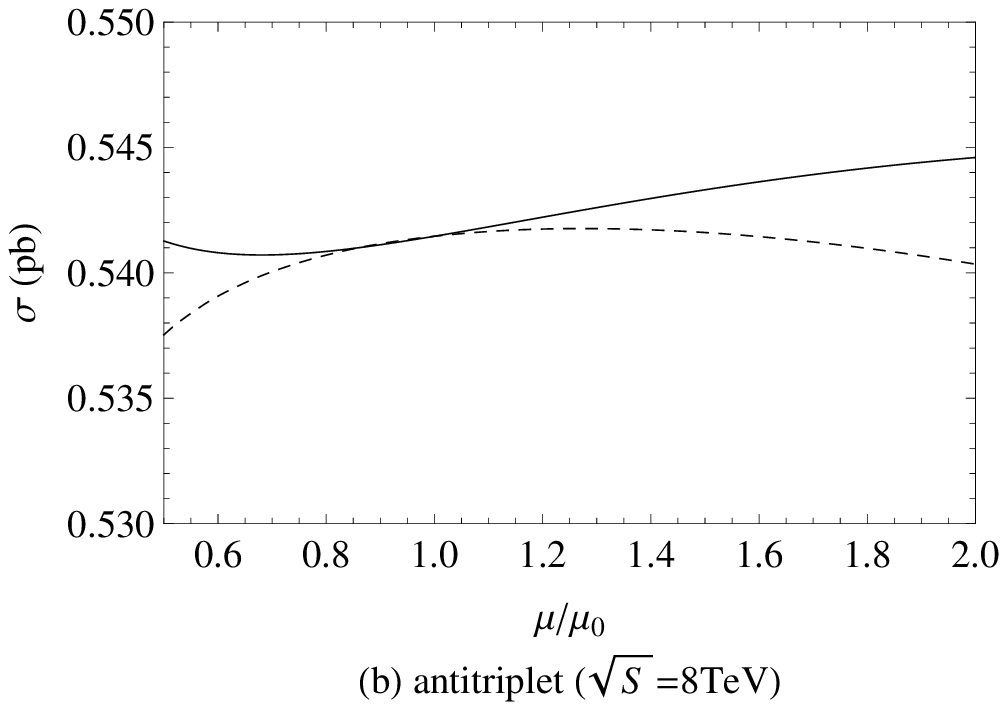}\\
  \includegraphics[scale=0.8]{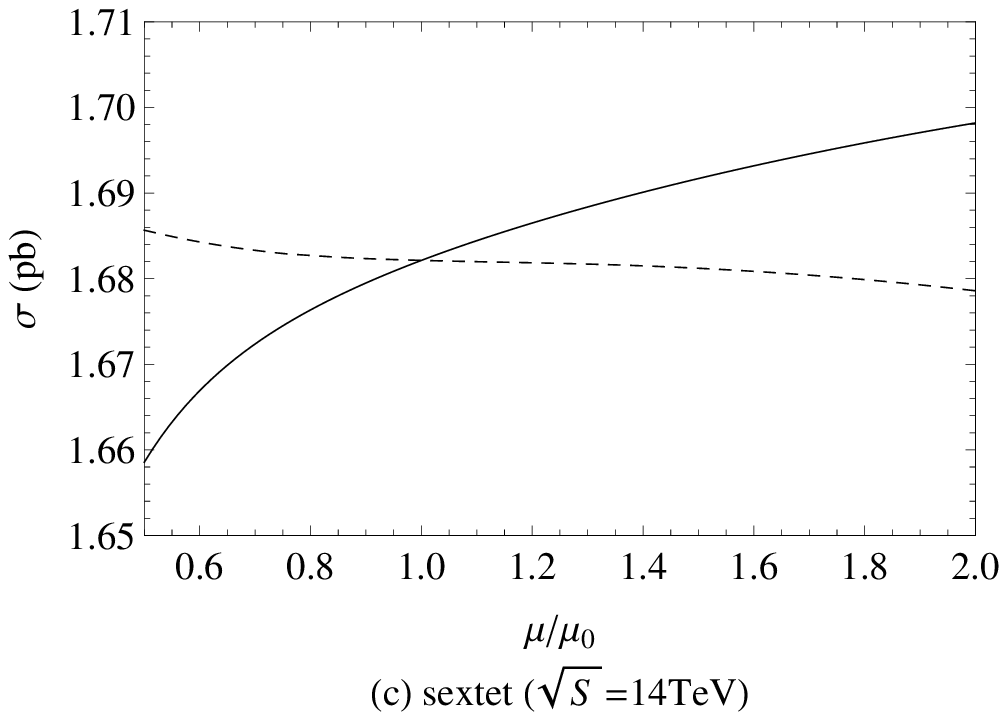}
  \includegraphics[scale=0.8]{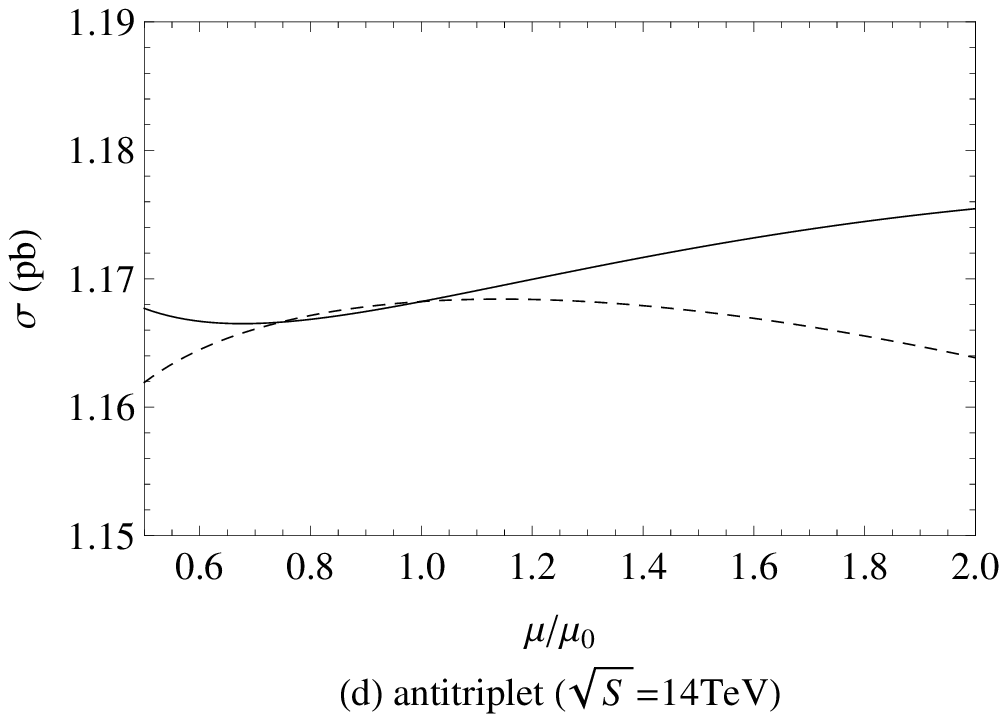}\\
  \caption{The $\mu_h$ and $\mu_s$ dependence of the resummed total cross sections. The solid and dashed lines represent $\mu_h$ and $\mu_s$ dependence, respectively. We set the scalar mass to be 1 TeV.}\label{fig.muh.mus.dep}
\end{figure}

Fig. \ref{fig.muh.mus.dep} shows the dependence of the resummed total cross section on $\mu_h$ and $\mu_s$. The scales are varied over the ranges $\mu_h^0/2<\mu_h<2\mu_h^0$ and $\mu_s^0/2<\mu_s<2\mu_s^0$, respectively. From Fig. \ref{fig.muh.mus.dep}, we can see that the $\mu_h$ dependence of the sextet is more sensitive than the antitriplet.

\begin{figure}[h]
  \centering
  \includegraphics[scale=0.8]{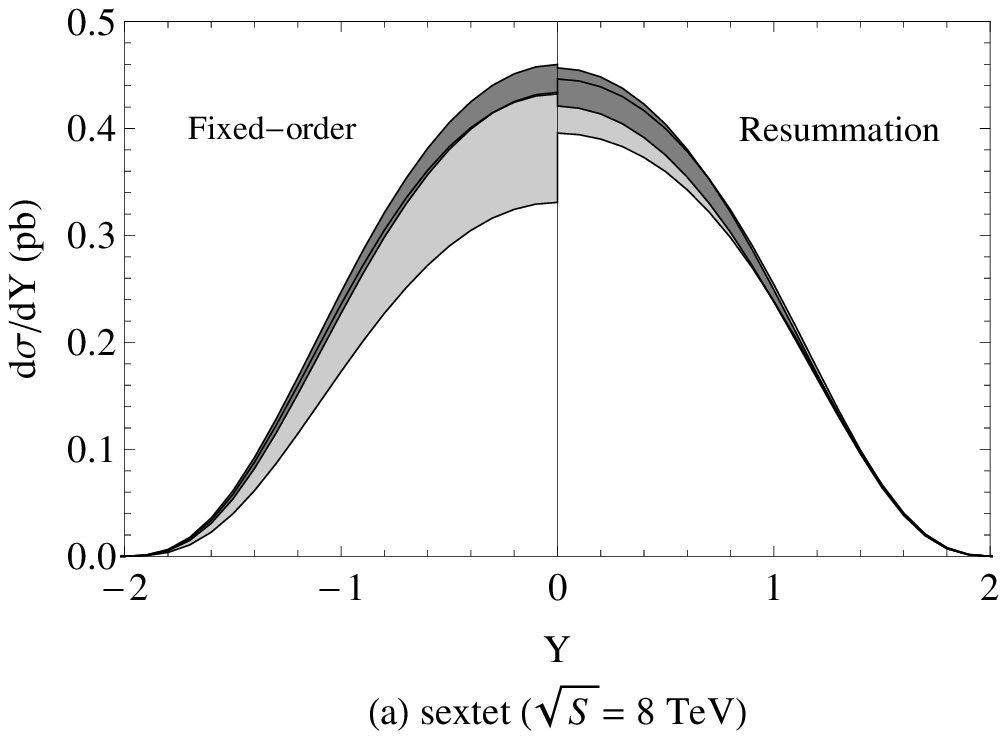}
  \includegraphics[scale=0.8]{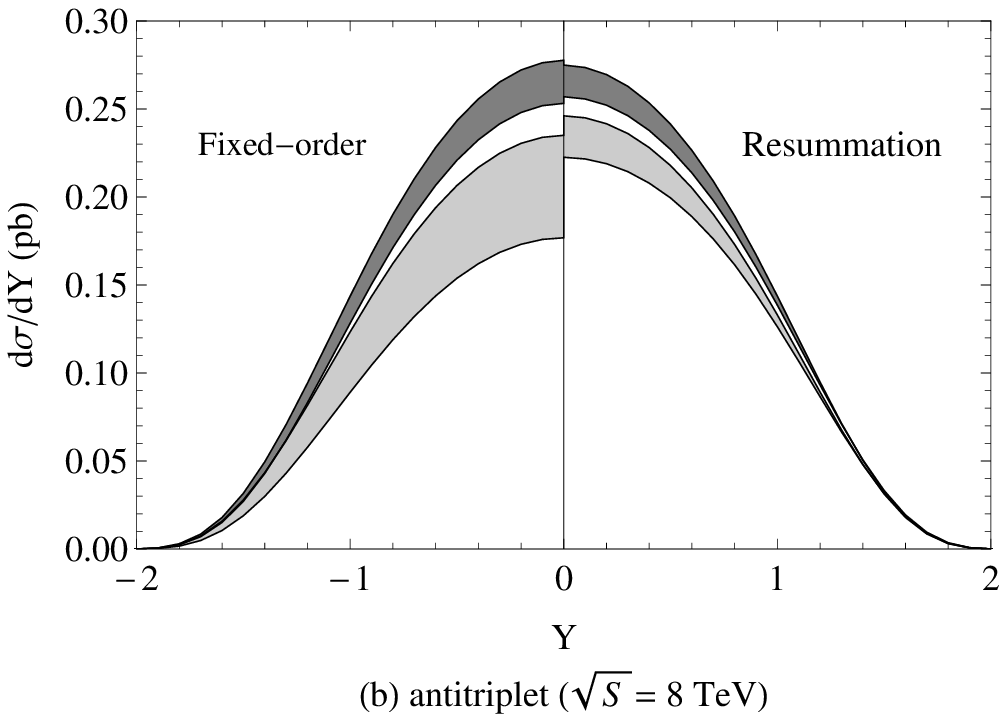}\\
  \includegraphics[scale=0.8]{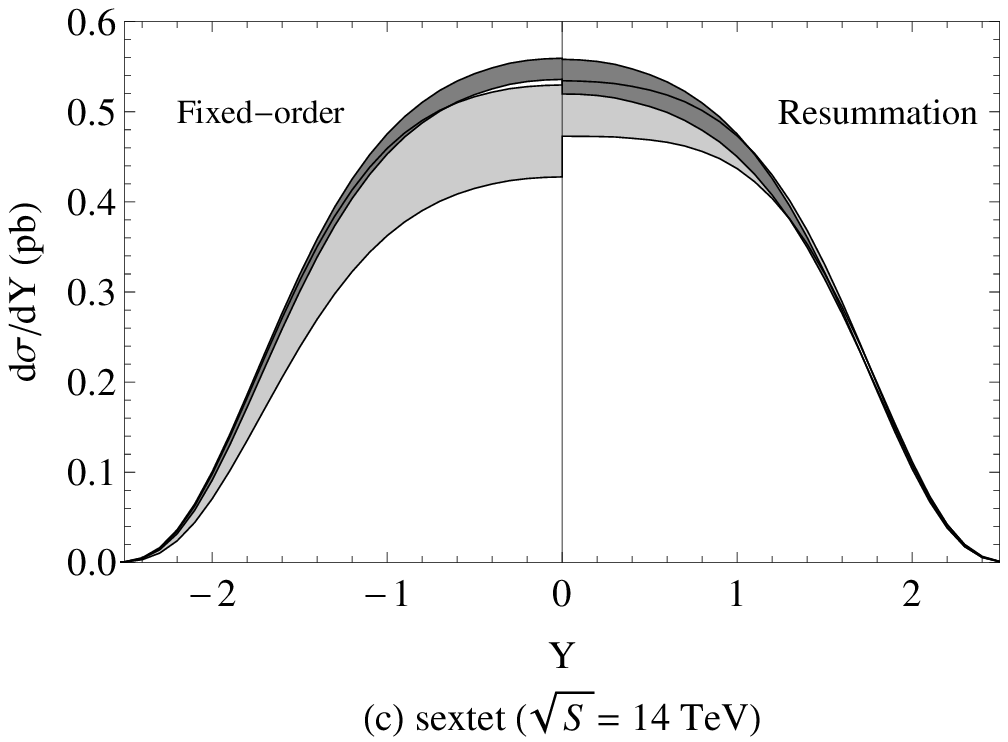}
  \includegraphics[scale=0.8]{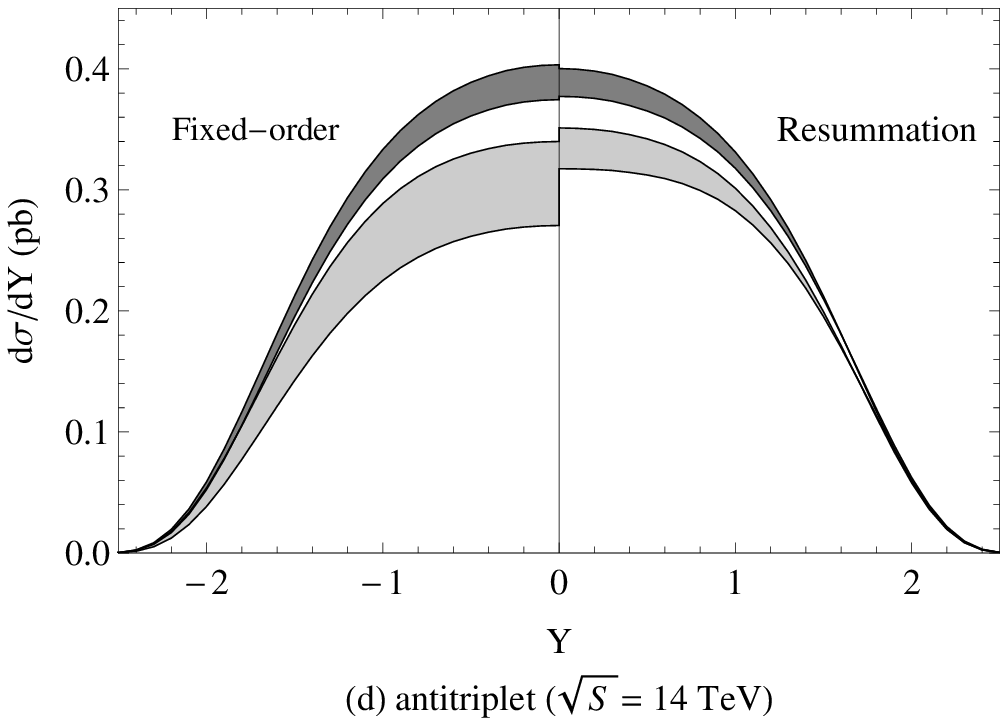}\\
  \caption{The comparison of the rapidity distributions between the combined resummation results and the fixed-order results for sextet and antitriplet. The scalar mass is set to be 1 TeV. The lighter bands stand for LO and NLL, while the darker represent NLO and NNLL$_{\text{approx}}$.}\label{fig.fixresdsdY}
\end{figure}

In Fig. \ref{fig.fixresdsdY}, we present the rapidity distributions, which compare the resummation results combined in Eq.(\ref{eq.combined resummation}) with the fixed-order results. The scale $\mu_f$ is varied over the range $m_{\phi}/2<\mu_f<2m_{\phi}$. We find that the shapes of the rapidity distribution of the resummation change slightly over the fixed-order results, and resummation reduces the scale dependence, except the NNLL$_{\text{approx}}$ results of the sextet cases. This is caused by the large color factor for the sextet ($C_D=10/3$ for the sextet, $C_D=4/3$ for the antitriplet). The terms containing a large color factor $C_D$, which is associated with the scale dependence of $\lambda$ and $\alpha_s$, will enlarge the scale dependence of the NNLL$_{\text{approx}}$ results of sextet.

\begin{figure}[h]
  \centering
  \includegraphics[scale=0.8]{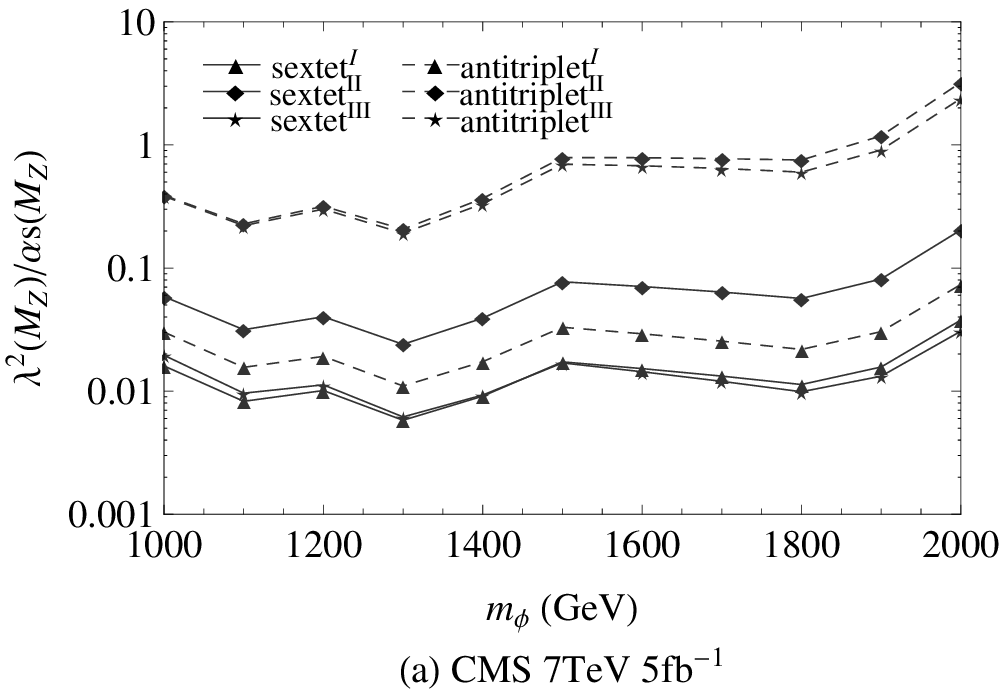}
  \includegraphics[scale=0.8]{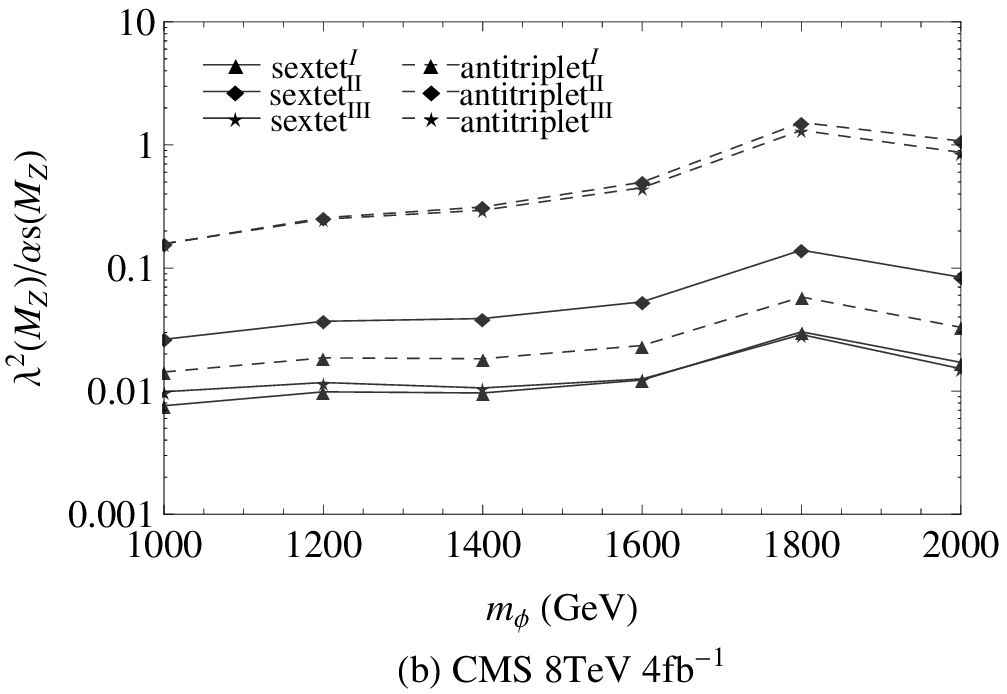}\\
  \includegraphics[scale=0.8]{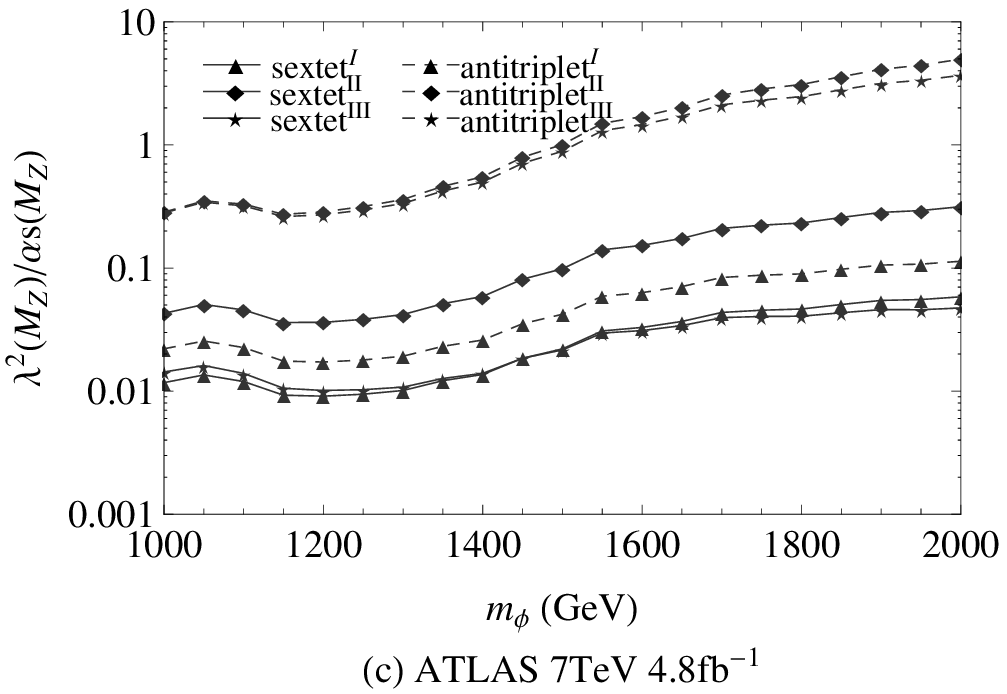}
  \includegraphics[scale=0.8]{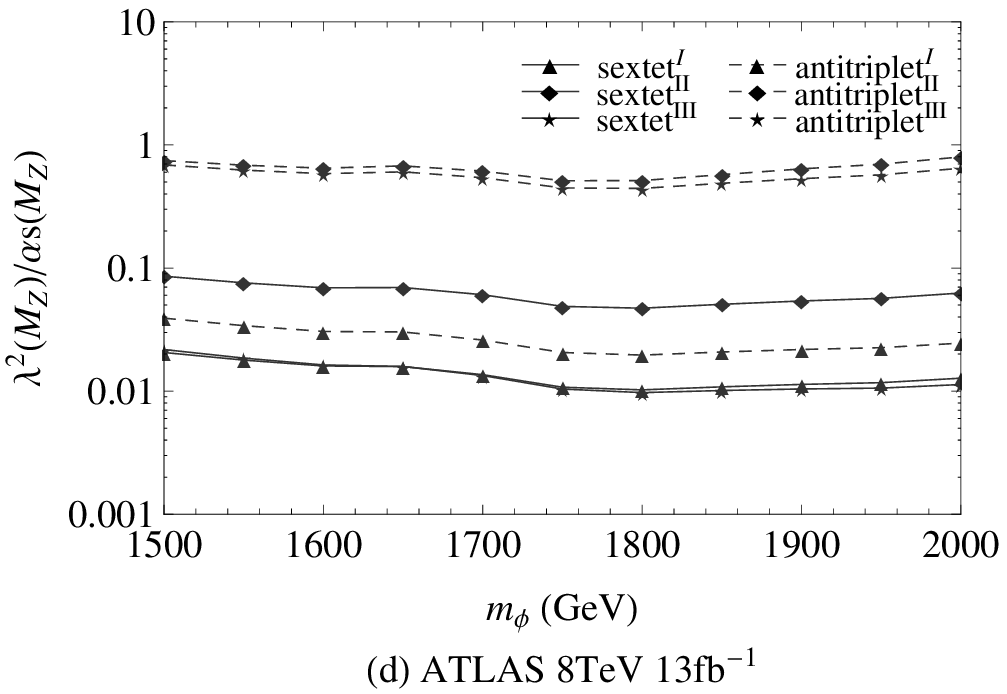}\\
  \caption{Constraint on the couplings $\lambda$ of the colored scalars with different electronic charges. }\label{fig.lambda.constraint}
\end{figure}

Finally, we use recent dijet data at the LHC to give constraints on the couplings $\lambda$. The CMS collaboration published the results of dijet production based on 5\ fb$^{-1}$ of 7 TeV data and 4\ fb$^{-1}$ of 8 TeV data \cite{CMS:2012yf,Chatrchyan:2013qha,CMS-PAS-EXO-12-059}, and the ATLAS collaboration based on 4.8\ fb$^{-1}$ of 7 TeV data  and 13\ fb$^{-1}$ of 8 TeV data \cite{ATLAS:2012pu,ATLAS-CONF-2012-148}.
Using the narrow-width-approximation \cite{Kauer:2007zc}, the total cross section can be written as
\begin{eqnarray}
\nonumber
  \sigma &=& \frac{(2\pi)^7}{2S}\int_{q_{min}^2}^{q_{max}^2}dq^2\int d\phi_p\ d\phi_d \big|\mathcal{M}_p(q^2)\big|^2
  \big[(q^2-m^2)^2+(m\Gamma)^2\big]^{-1}\big|\mathcal{M}_d(q^2)\big|^2 \\
   &=& \frac{(2\pi)^8}{4Sm\Gamma}\int d\phi_p\big|\mathcal{M}_p(q^2)\big|^2\int d\phi_d\big|\mathcal{M}_d(q^2)\big|^2.
\end{eqnarray}
After fitting the dijet data, we can give the constraints on the couplings. Since there is no direct theoretical requirement on the couplings between the colored scalars and different quarks, we use a common value for the coupling $\lambda$ here. The colored scalars with different electronic charges couple to different quarks, and then they receive different constraints.
In Fig. \ref{fig.lambda.constraint}, we show the results of the constraints on the couplings. The most stringent constraint on $sextet^{I}$ is $\lambda^2(M_Z)\geq0.006\alpha_s(M_Z)$, and similarly the other constraints are 0.024$\alpha_s(M_Z)$, 0.006$\alpha_s(M_Z)$, 0.011$\alpha_s(M_Z)$, 0.16$\alpha_s(M_Z)$ and 0.16$\alpha_s(M_Z)$ for $sextet^{II}$, $sextet^{III}$, $antitriplet^{I}$, $antitriplet^{II}$ and $antitriplet^{III}$, respectively.

\section{Conclusion}
\label{sec:conclusion}

We have studied the threshold resummation effects in the single production of the color sextet (antitriplet) scalars at the LHC with the soft-collinear effective theory. We find that the resummation effects increase the NLO total cross section by about $2\%$ and $0.2\%$ for 1 TeV color antitriplet and sextet scalar, respectively, and $5\%$ and $3\%$ for 2 TeV color antitriplet and sextet scalar, respectively, at the 8 TeV LHC. The resummation effects improve the scale dependence of the cross section and the rapidity distribution generally. But in the case of the rapidity distribution of the color sextet scalar, the scale dependence is not improved because of the large color factor $C_D$ ($C_D=10/3$ for the sextet, $C_D=4/3$ for the antitriplet) enlarging the scale dependence. Besides, we use recent dijet data from the LHC to give constraints on the couplings. For different colored scalars with different electronic charges, the most stringent constraints of $\lambda^2(M_Z)$ range from 0.006$\alpha_s(M_Z)$ to 0.16$\alpha_s(M_Z)$.

\begin{acknowledgments}
We would like to thank Hua Xing Zhu, Jian Wang and Qing Hong Cao for useful discussions. This work was supported in part by the National Natural
Science Foundation of China, under Grants No. 11375013 and No. 11135003.
\end{acknowledgments}



\appendix

\section{Relevant Feynman Diagrams}
\label{sec:feynmandiagrams}

\begin{figure}[h]
  \centering
  \includegraphics[scale=1]{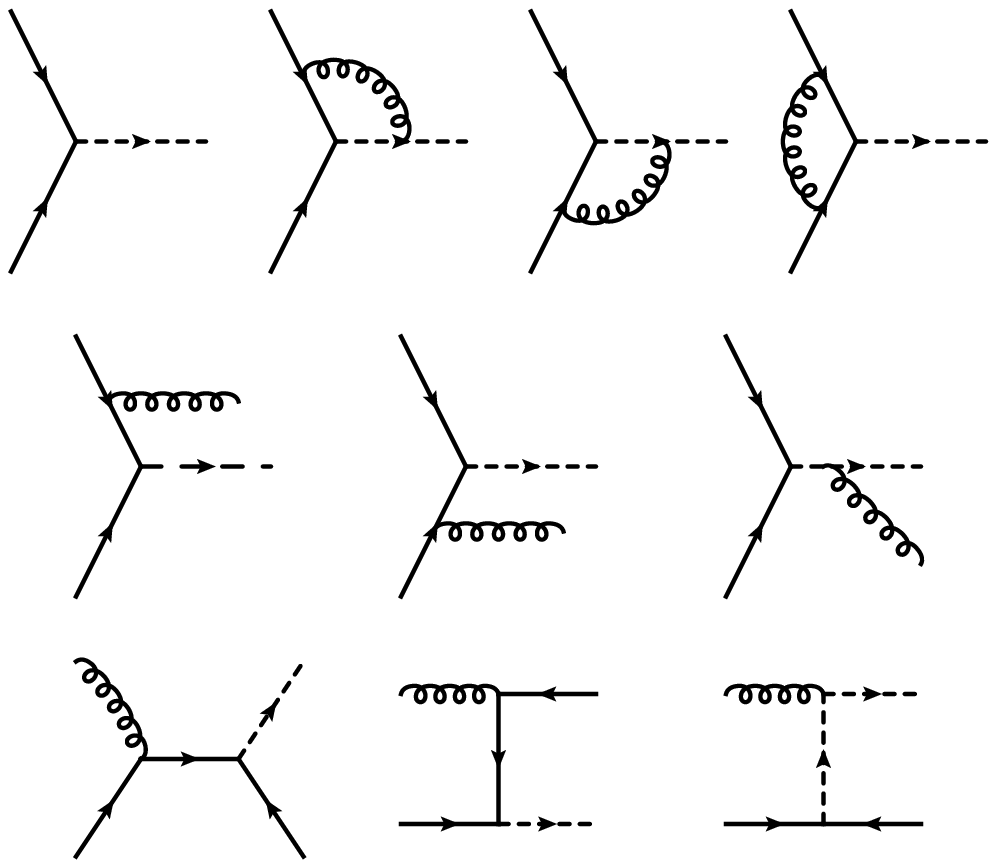}\\
  \caption{Relevant Feynman diagrams for the production of the colored scalar.}\label{fig.fd}
\end{figure}

Relevant Feynman diagrams for the production of the colored scalar are shown in Fig. \ref{fig.fd}.

\bibliography{CSSr1}{}

\end{document}